\documentclass[aps,twocolumn,showpacs,preprintnumbers,amsmath,amssymb,nofootinbib,showkeys]{revtex4-1}

\usepackage{epsfig,graphicx,amsmath,amssymb}%,citesort}
\usepackage{picture}
\usepackage{braket}
\usepackage{slashed}
\usepackage{changes}
\usepackage{soul}
\usepackage{ulem}
\usepackage{dsfont}
\usepackage{xcolor} \definecolor{orange}{rgb}{1,0.5,0} \definecolor{bluegreen}{rgb}{0,0.4,0.4} %\definecolor{bluegreen}{rgb}{0.23,0.70,0.34}
\usepackage{marginnote}
\usepackage{relsize}
\def\babar{\mbox{\slshape B\kern-0.1em{\smaller A}\kern-0.1emB\kern-0.1em{\smaller A\kern-0.2em R}}}

\begin{document}

\title{Pseudoscalar-pole contribution to the $(g_{\mu}-2)$: a rational approach}

\author{Pere Masjuan} \email{masjuan@ifae.es}
\affiliation{Grup de F\'{\i}sica Te\`orica, Departament de F\'{\i}sica, 
  Universitat Aut\`onoma de Barcelona,  and Institut de F\'{\i}sica d'Altes Energies (IFAE), The Barcelona Institute of Science and Technology (BIST), 
  Campus UAB, E-08193 Bellaterra (Barcelona), Spain}
\author{Pablo Sanchez-Puertas} \email{sanchezp@ipnp.troja.mff.cuni.cz}
\affiliation{Faculty of Mathematics and Physics, Institute of Partice and Nucelar Physics, Charles University in Prague, V Hole\v{s}ovi\v{c}k\'ach 2, Praha 8, Czech Republic}
\begin{abstract}
We employ a mathematical framework based on rational approximants in order to calculate the pseudoscalar-pole piece of the hadronic light-by-light contribution to the anomalous magnetic moment of the muon, $a_{\mu}^{\textrm{HLbL};P}$. The method is systematic and data based, profiting from over 13 different collaborations, and able to ascribe, for the first time, a systematic uncertainty which provides for the model independence. As a result, we obtain $a_{\mu}^{\textrm{HLbL};P}=94.3(5.3) \times10^{-11}$, which uncertainty is well below the one foreseen at future experiments measuring the $(g_{\mu}-2)$. 
\end{abstract}


\maketitle

\section{\label{sec:intro}Introduction}

The anomalous magnetic moment of the muon, $a_{\mu}\equiv(g_{\mu}-2)/2$, represents one of our finest tests of the Standard Model (SM) of particle physics and its most recent measurement, $a_{\mu}^{\textrm{exp}}=116592091(63)\times10^{-11}$~\cite{Bennett:2006fi,Agashe:2014kda}, has reached the astonishing precision of $0.54$~ppm. At such precision $a_{\mu}$ does not only provide a beautiful test of our understanding of elementary interactions, but represents an interesting probe of physics beyond the SM. Indeed, there is at present a discrepancy among experiment and theory of around $3\sigma$~\cite{Jegerlehner:2009ry}\footnote{The progress on this field is captured in at least three recent dedicated workshops on the $(g_{\mu}-2)$  \cite{Czyz:2013zga,Benayoun:2014tra,Proceedings:2016bng}.
}. For this reason, two new experiments have been projected both at Fermilab~\cite{LeeRoberts:2011zz} and J-PARC~\cite{Mibe:2010zz}, which expect to measure $a_{\mu}$ at a precision of around $0.14$~ppm and would shed light on the nature of the present discrepancy. However, such a tremendous effort on reducing the experimental uncertainty would be in vain unless the current theoretical calculations would reach a similar accuracy. 
At present, the theoretical error is dominated by two different hadronic contributions: the hadronic vacuum polarization (HVP), which amounts to $58$ ppm to $a_{\mu}$ and with an uncertainty of around $0.36$~ppm~\cite{Jegerlehner:2009ry}, and the hadronic light-by-light (HLbL), which amounts to $0.87$ ppm to $a_{\mu}$ and with an uncertainty of around $0.33$~ppm~\cite{Jegerlehner:2009ry}, leading to a total theoretical error reading $0.49$ ppm. These calculations, involving complicated loop integrals, are hindered via the non-perturbative hadronic physics dominating the loop integrals. Fortunately, such complications can be overcome for the dominant contribution, the HVP,  since it is related through the optical theorem to the $\sigma(e^+e^-\to\textrm{hadrons})$ cross section. It is expected that, in the near future, the ongoing experimental program will allow to reduce the HVP errors according to what future experiments require. Still, such effort would be fruitless unless a similar reduction in the precision of the HLbL is achieved, which would otherwise dominate the theoretical SM uncertainty and  make the experimental efforts pointless. 

By contrast to the  HVP, the HLbL entails a much richer structure that avoids an easy connection to data. As a consequence, the existing calculations have required certain modeling and approximation procedures. Ascribing them a systematic error is a difficult task, but the variety of the present results~\cite{Jegerlehner:2009ry} already suggests an error which is potentially larger than future experiments' precision, which demands a new, more accurate and less model-dependent evaluation. Among the different contributions to the hadronic-light-by-light, the pseudoscalar-pole seems to dominate the full quantity, requiring therefore the best precision. Fortunately, such quantity can be rigorously defined in a quantum field theory and related to the pseudoscalar transition form factors (TFFs), which are observable quantities. This offers an opportunity to perform a data-driven approach for this contribution provided that a reliable method is established. In this work, we discuss a novel method based on Canterbury approximants (bivariate Pad\'e approximants) which provides a mathematically and data-based description for the involved TFFs in the space-like (SL) region, allowing then for a model-independent calculation for the pseudoscalar-pole contribution to the hadronic-light-by-light which includes, for the first time, the sought systematic error which has been missing in previous calculations. Special attention on the reasons which justify the use of Canterbury approximants in contrast to previous resonance approaches is given.

The paper is structured as follows: first, we briefly introduce the most general HLbL contribution to $a_{\mu}$ and its general features in Sect.~\ref{sec:hlbl}. Its main piece, the pseudoscalar pole, is presented in Sect.~\ref{sec:ppole}; this includes also a brief overview of current theoretical approaches and motivates the reasons for our new study based on Canterbury approximants. These are subsequently presented in Sect.~\ref{sec:ca}. Finally, we give our  results in Sect.~\ref{sec:results} and discuss the role of future data in Sect.~\ref{sec:data}. 
Much information is relegated to the appendices, including, among others, the impact of $P \to \bar{\ell}\ell$ decays, a discussion concerning the light-quark TFF, comments concerning the pseudoscalar-exchange approach, a discussion on dispersion relations, and our most-updated data input profitting from over 13 different collaborations.

\section{\label{sec:hlbl}Hadronic light-by-light contribution}

\begin{figure}[b]
    \centering
    \includegraphics[width=0.55\linewidth]{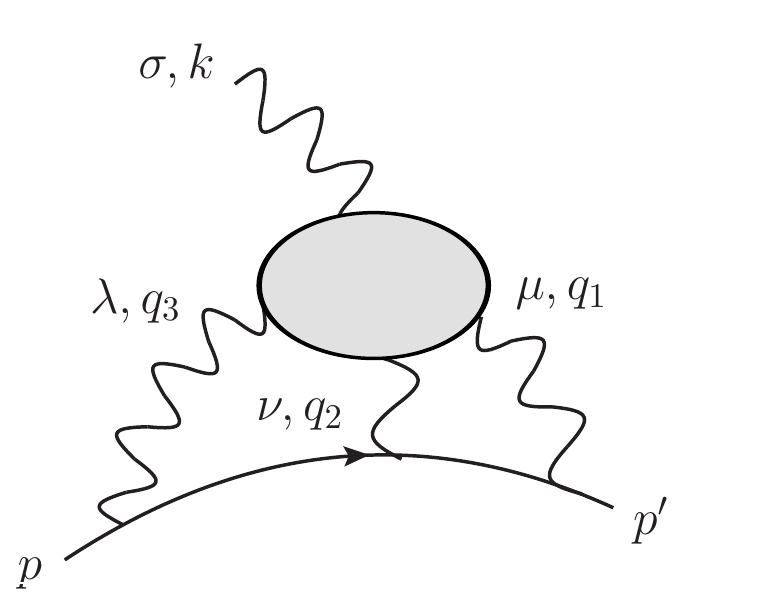}
    \caption{HLbL contribution to $a_{\mu}$. The grey blob represents the HLbL tensor. $q_{1-3}$ are outgoing momenta, whereas $k$ is the incoming external photon momentum.}
    \label{fig:HLbL}
\end{figure}

The HLbL contribution to $a_{\mu}$ (cf. Refs.~\cite{Jegerlehner:2009ry,Prades:2009tw}) is depicted in Fig.~\ref{fig:HLbL}, where the gray blob represents the HLbL tensor defined as

\begin{align}
\Pi_{\mu\nu\lambda\rho}(q_1,q_2,q_3) \! & = \! \int d^4x \int d^4y \int d^4z e^{i(q_1\cdot x  +  q_2\cdot y   +   q_3\cdot z)}  \nonumber \\
&  \times   \bra{\Omega} T \{ j_{\mu}(x)j_{\nu}(y)j_{\lambda}(z)j_{\rho}(0) \} \ket{\Omega}, \label{eq:hlbltensor}
\end{align}
with all the momenta, $q_i$, outgoing. The resulting contribution to $a_{\mu}$ can be expressed in terms of this hadronic quantity using projection techniques that allow to specialize to the kinematical limit relevant to $a_{\mu}$ $(k\to0)$ in advance, obtaining~\cite{Knecht:2001qf}:
\begin{widetext}
\begin{align}
a_{\mu}^{\textrm{HLbL}} = 
 &  \frac{-ie^6}{48m_{\mu}}  \int  \frac{d^4q_1}{(2\pi)^4}  \int  \frac{d^4q_2}{(2\pi)^4}  \frac{1}{q_1^2 q_2^2 (q_1+q_2)^2} \frac{1}{(p-q_1-q_2)^2 -m_{\mu}^2} \frac{1}{(p-q_1)^2 -m_{\mu}^2} \times \!  \textrm{tr}\Big( (\slashed{p}+m_{\mu})[\gamma^{\rho},\gamma^{\sigma}]  \nonumber \\
&  (\slashed{p}+m_{\mu})  \gamma^{\mu}(\slashed{p}- \slashed{q}_1 \! +m_{\mu})   \gamma^{\nu}(\slashed{p}- \slashed{q}_1 \! - \slashed{q}_2 \! +m_{\mu})\gamma^{\lambda} \Big) \!
 \left[\frac{\partial}{\partial k^{\rho}}\Pi_{\mu\nu\lambda\sigma}(q_1,q_2,k-q_1-q_2) \right]_{k\rightarrow 0}.  \label{eq:HLbLint}
\end{align}
\end{widetext}
Such quantity requires as input the HLbL tensor at all energy scales, including the (non-perturbative) low energies, which turn out to play the major role in numerical calculations as we will illustrate for the particular case of the $\pi^0$-pole contribution. Qualitatively, this can be understood, after Wick rotation, from the propagators in Eq.~\eqref{eq:HLbLint}. 

A full description of the HLbL tensor is far from a trivial task. As an illustration, while the dominant HVP requires a scalar function depending on a single energy scale,  the HLbL tensor involves 138 scalar functions\footnote{That number is however reduced after considering the Ward identities and the kinematic configuration relevant to $a_{\mu}$~\cite{Bijnens:1995xf,Colangelo:2015ama}.} and six scalar variables (the four photon virtualities, $\{q_{i}^2\}_{i=1-4}$, and 2 Mandelstam variables).
In order to deal with such object, E.~de~Rafael proposed more than twenty years ago~\cite{deRafael:1993za} to split the most relevant contributions to this tensor according to a combined expansion in terms of the chiral and the large-number-of-colors ($N_c$) limits of QCD; whereas large-$N_c$ represents the only known perturbative approach to QCD, the chiral expansion, in powers of momentum ($q_i^2$), helps to identify those contributions which play the main role at low energies --- the most important in the calculation. According to this proposal, the leading contributions to the HLbL tensor are the pseudoscalar loop contributions, at order $\mathcal{O}(N_c^0,q^4)$, and the pseudoscalar-pole terms, at order $\mathcal{O}(N_c,q^6)$. Following these ideas, subleading contributions will account for heavier resonances and the continuum quark-loop contributions, all of them of order $\mathcal{O}(N_c,q^8)$. It remains then the task to calculate all the relevant contributions as accurate and precise as possible. 

Actually, most of the results in the literature follow de Rafael's proposal (see Refs.~\cite{Hayakawa:1997rq,Bijnens:2001cq,Knecht:2001qg,Blokland:2001pb,Melnikov:2003xd,Dorokhov:2008pw,Nyffeler:2009tw,Jegerlehner:2009ry,Prades:2009tw,Hong:2009zw,Cappiello:2010uy,Kampf:2011ty,Masjuan:2012wy,Bijnens:2012an,Escribano:2013kba,Roig:2014uja,Dorokhov:2015psa}, including full and partial contributions to $a_{\mu}^{\textrm{HLbL}}$) finding values for $a_{\mu}^{\textrm{HLbL}}$ between basically $6 \times 10^{-10}$ and up to almost $14 \times 10^{-10}$. Among them, the Jegerlehner and Nyffeler review~\cite{Jegerlehner:2009ry},  quoting $(11.4 \pm 4.0) \times 10^{-10}$, and the \textit{Glasgow consensus}~\cite{Prades:2009tw}, written by Prades, de Rafael, and Vainshtein, and quoting $(10.5 \pm 2.6) \times 10^{-10}$,  represent in our opinion the standard reference values for the HLbL. They agree well since they only differ by few subtleties: they both used the model from Knecht and Nyffeler~\cite{Knecht:2001qf} to account for the dominant contribution, but differ on how to implement the high-energy QCD constrains and the error propagation. Neither of both approaches contain however systematic errors from the chiral and large-Nc limits~\cite{Masjuan:2012wy,Masjuan:2012qn,Masjuan:2007ay,Masjuan:2008fr}, which are difficult to estimate (cf. also the discussion in Ref.~\cite{Masjuan:2014rea}).  An attempt to consider a dispersion relation to calculate a subset of the pieces appearing in de Rafael's proposal has been recently proposed in Refs.~\cite{Colangelo:2014dfa,Pauk:2014rfa}, but numerics are still to come.  All in all, even though the QCD features for the HLbL are well understood~\cite{Jegerlehner:2009ry,Prades:2009tw}, the details of the particular calculations are important to get the numerical result to the final required precision.

Alternative approaches to calculate the HLbL contribution exist as well. As an example, a ballpark estimate can be obtained using the analytical result for the heavy quark loop contribution to the HLbL in the lines of Refs.~\cite{Pivovarov:2001mw,Erler:2006vu,Boughezal:2011vw,Greynat:2012ww,Masjuan:2012qn}, finding somewhat higher values around $(12\div 17)  \times 10^{-10}$. Likewise, the 
Dyson-Schwinger approach adopted in Ref.~\cite{Fischer:2010iz,Goecke:2010if,Goecke:2012qm} and the lattice QCD simulations in Refs.~\cite{Blum:2014oka,Blum:2015gfa,Blum:2016lnc,Green:2015sra} are also approaches which do not follow the same procedure. Whereas yet incomplete and with some progress still required, promising advances have been reported already~\cite{Goecke:2010if,Eichmann:2014ooa,Blum:2014oka,Blum:2015gfa,Blum:2016lnc}.

Despite the combined chiral and large-$N_c$ counting yields the pseudoscalar loop contributions and the pseudoscalar-pole terms as the leading contributions to the HLbL tensor, suggesting a similar size, the situation is more subtle. The more careful discussion in Ref.~\cite{Melnikov:2003xd} observed that the typical size of the momentum running in the pion loop turns out to be of order $4m_{\pi}$~\cite{Melnikov:2003xd}, which implies a slow convergence of the chiral expansion with non-leading terms logarithmically enhanced. This reduces the pion loop contribution with respect to the pion pole.  Phenomenologically, the former is found to be around four times smaller than the latter \cite{Jegerlehner:2009ry}, which becomes thereby the most relevant contribution to be calculated among the different HLbL contributions. Since such contribution is typically found to be of order $10^{-9}$, in order to meet the $0.14$~ppm precision of future experiments, a precision below $10\%$ is a priori desired, which is beyond traditional approaches' performance\footnote{As we illustrate in the following sections, the current values~\cite{Jegerlehner:2009ry,Prades:2009tw} entail potential large systematic errors above $10\%$.}. 

In the following section we outline concisely what this contribution refers to and its relation to the pseudoscalar TFFs. In addition, the relevant kinematical regions of interest are identified, whereby the requirements that a TFF parameterization necessitates are obtained, which will naturally motivate our approach employing Canterbury approximants.

\section{Pseudoscalar-pole contribution\label{sec:ppole}}

\begin{figure}[t]
 \includegraphics[width=\linewidth]{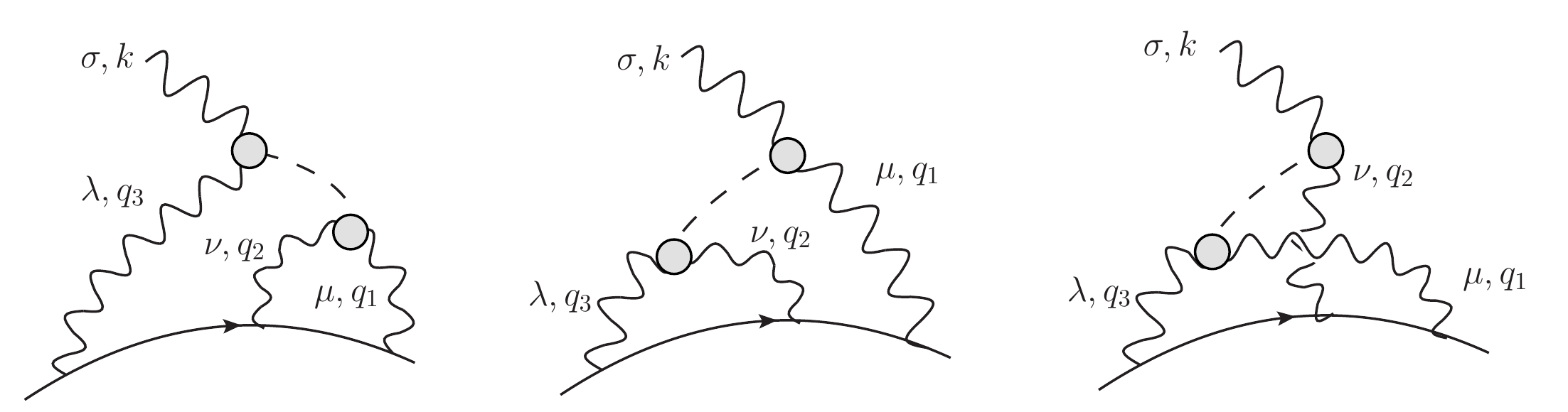}
 \caption{The pseudoscalar-pole contribution to $a_{\mu}^{\textrm{HLbL}}$.}
 \label{fig:pspole}
\end{figure}

The pseudoscalar-pole contribution to the HLbL tensor can be easily obtained within the language of Green's functions. Inserting the \textit{identity} as a sum over the QCD spectrum ($\mathds{1} = \sum_X\int d\Pi_X\ket{X}\bra{X}$, with $\ket{X}$ on-shell intermediate hadronic states) within the HLbL tensor Eq.~\eqref{eq:hlbltensor}, it is obtained that such function exhibits well-isolated poles for the lightest pseudoscalar states $P=\{\pi^0,\eta,\eta'\} \in X$ which contribution to $a_{\mu}^{\textrm{HLbL}}$ is depicted in Fig.~\ref{fig:pspole}. This is, the HLbL tensor can be expressed as (sum over $P$ assumed)
\begin{multline}
\Pi^{\mu\nu\lambda\rho}(q_1,q_2,q_3)   = \!  \int\! d^4x \! \int\! d^4z e^{iq_1\cdot x}e^{iq_3\cdot z} \frac{i}{q^2-m_P^2+i\epsilon}  \\ 
 \times \! \bra{0} T\{ j^{\mu}(x) j^{\nu}(0) \} \!\ket{P} \bra{P} T\{ j^{\lambda}(z) j^{\rho}(0) \} \!\ket{0} + \textrm{OT}, \label{eq:GreenToTFF}
\end{multline}
where $q=q_1+q_2=k-q_3$ and with OT referring both, to different  time orderings (crossed $t$ and $u$ channels) and to additional contributions from the QCD spectrum ($X\neq P$ and not necessarily of pseudoscalar nature) which do not become singular as $q^2\to m_P^2$. Besides, for the on-shell pseudoscalar states, the matrix elements defined above are related to the pseudoscalar TFFs (gray blobs in Fig.~\ref{fig:pspole}), defined as\footnote{Note that the $(ie)^2$ coupling is already implicit in Eq.~\eqref{eq:HLbLint}.} 
\begin{align}\nonumber
i\mathcal{M}^{\mu\nu}_{P\rightarrow\gamma^*\gamma^*} &\equiv
\int d^4x e^{iq_1\cdot x} \bra{\Omega} T\{j^{\mu}(x)j^{\nu}(0) \} \ket{P} \\ 
&\equiv -i\epsilon^{\mu\nu\rho\sigma}q_{1\rho}q_{2\sigma}F_{P\gamma^*\gamma^*}(q_1^2,q_2^2), \label{eq:TFF}
\end{align}
with $\epsilon^{0123}=+1$. As a consequence, the contribution from the pseudoscalar poles can be calculated as model-independent as the employed TFF description is, without incurring in any ambiguity.
OT in Eq.~\eqref{eq:GreenToTFF} defines heavier states --- which could be hardly included in this way given their widths~\cite{Agashe:2014kda} --- and continuum contributions. These processes cannot however induce a non-analytic behavior at energies close to the lightest pseudoscalar masses, which allow to disentangle the pseudoscalar-pole contribution, as previously said, unambiguously.

At this point, it is worth to make a brief digression and comment on what is referred as \textit{the pseudoscalar-exchange} contribution, an analytically similar but theoretically different contribution to the pseudoscalar-pole one. The authors from Ref.~\cite{Melnikov:2003xd} realized that the pseudoscalar-pole contribution could not reproduce the high-energy QCD constraints imposed by the OPE for the {\emph{full}} HLbL tensor. However, by setting the TFF involving the external photon to a constant one, such constraint can be satisfied. Their approach, as they mention~\cite{Melnikov:2003xd}, was meant as a model to interpolate the full HLbL tensor from the low to the high energies, which includes, effectively, excited pseudoscalar resonances. Later on, Ref.~\cite{Jegerlehner:2009ry} criticized such approach and introduced what is known as the \textit{off-shell} TFF. This approach intends to consider all the pseudoscalar contributions (i.e. heavier resonances and continuum) into an off-shell pseudoscalar TFF by connecting with the $\langle VVP \rangle$ Green's function and imposing its well-known high-energy behavior~\cite{Melnikov:2003xd}. We note that such procedure cannot be rigorously derived as a pole-contribution, and could be thought as a model interpolating the low- and high-energy behavior for the exchange of pseudoscalar-like resonances in Eq.~\eqref{eq:GreenToTFF}. 

It is not our intention to discuss how the high energies of the HLbL tensor should be implemented in terms of the pseudoscalar-pole contribution; our more modest concern is to discuss a model-independent data-based description for the pseudoscalar-pole contribution to the $a_{\mu}^{\textrm{HLbL}}$, inspired by new and forthcoming experimental results, lattice simulations, and dispersive representations.

\begin{figure*}[t]
\centering
   \includegraphics[width=0.3\linewidth]{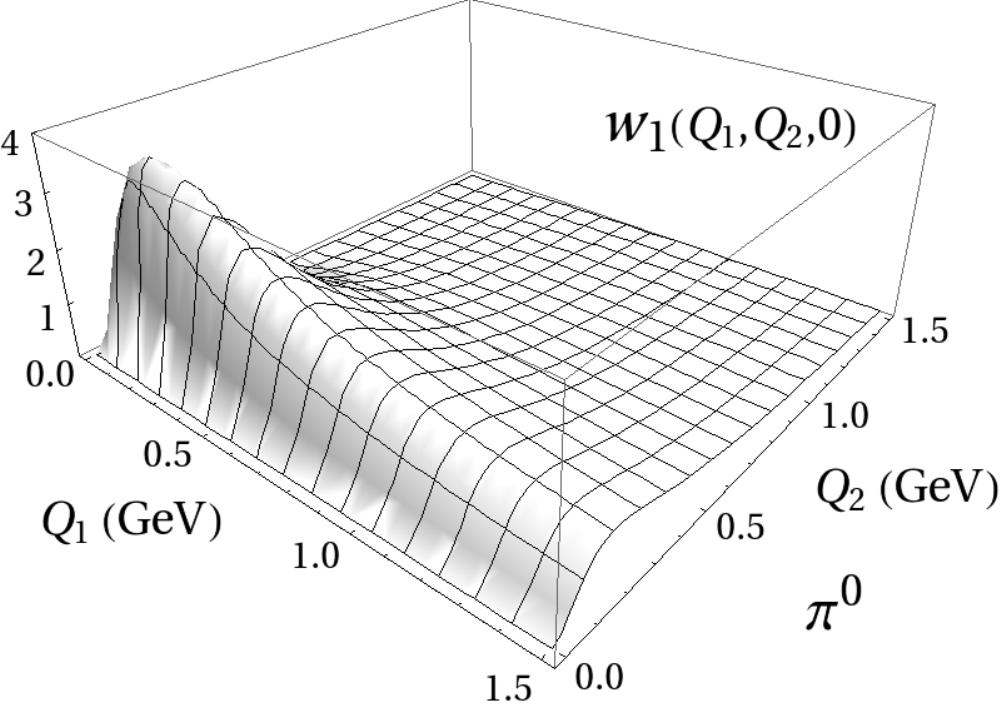}
   \includegraphics[width=0.3\linewidth]{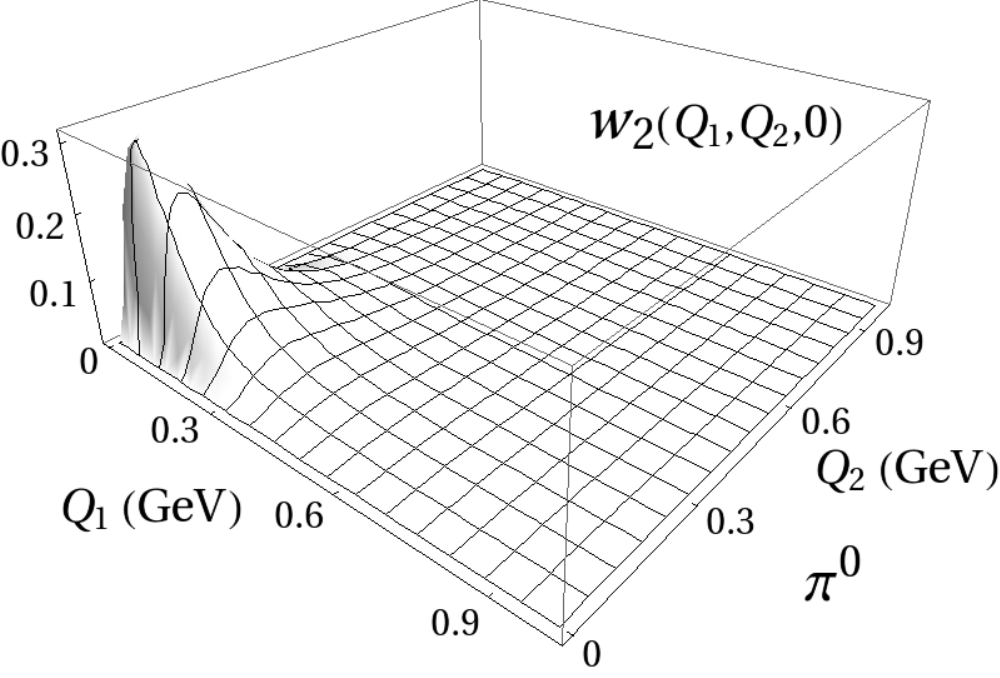}
   \includegraphics[width=0.3\linewidth]{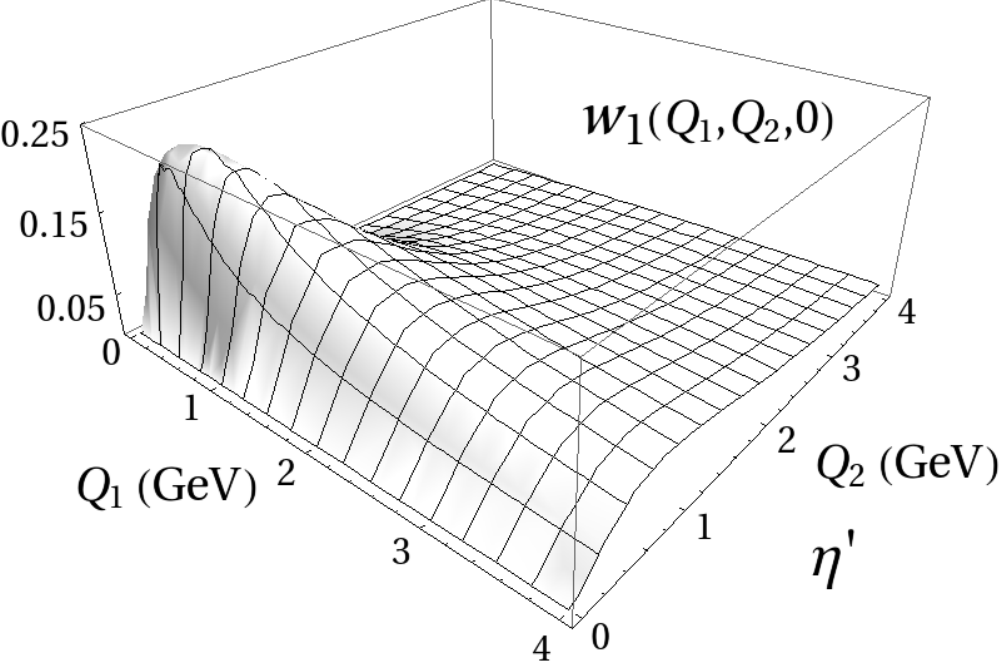}
\caption{The $w_{1,2}(Q_1,Q_2,t)$ integrands in Eq.~\eqref{eq:hlblpex} for $t=0$ and a constant TFF. The first two stand for $w_1(Q_1,Q_2,t)$ and $w_2(Q_1,Q_2,t)$ functions for the $\pi^0$ case; the third one stands for $w_1(Q_1,Q_2,t)$ for the $\eta'$ case. Note the difference in scales.}
\label{fig:kernel}
\end{figure*}

The pseudoscalar-pole contribution to the HLbL tensor (Eq.~\eqref{eq:GreenToTFF}) is given explicitly in Eq.~\eqref{eq:polehlbl} and involves the $s,t$ and $u$ channels as illustrated in Fig.~\ref{fig:pspole}. Inserting this into Eq.~\eqref{eq:HLbLint}, performing the Wick rotation and using the Gegenbauer technique to perform angular integrations~\cite{Knecht:2001qf}, the pseudoscalar-pole contribution to $a_{\mu}^{\textrm{HLbL}}$ can be expressed as an integral over one angular variable and two space-like momenta~\cite{Jegerlehner:2009ry}:
\begin{align}
a_{\mu}^{\textrm{HLbL};P} \!= &   \frac{-2\pi}{3}\left(\frac{\alpha}{\pi}\right)^3 \int_0^{\infty}dQ_1dQ_2 \int_{-1}^{+1}dt \sqrt{1-t^2} Q_1^3Q_2^3  \nonumber \\
                            & \ \times  \left[  \frac{F_1 I_1(Q_1,Q_2,t)}{Q_2^2+m_{P}^2}  +    \frac{F_2 I_2(Q_1,Q_2,t)}{Q_3^2+m_{P}^2}  \right] \nonumber \\
                            \equiv & \left(\frac{\alpha}{\pi}\right)^3 \int_0^{\infty} \! \! dQ_1dQ_2 \int_{-1}^{+1} \! \!  \! \! dt  \left[ w_1F_1  +  w_2 F_2  \right], \label{eq:hlblpex}
\end{align}
where $F_1$ and $F_2$ are defined as
\begin{align} \nonumber
F_1 &= F_{P\gamma^*\gamma^*}(Q_1^2,Q_3^2)F_{P\gamma^*\gamma}(Q_2^2,0), \\ F_2 &= F_{P\gamma^*\gamma^*}(Q_1^2,Q_2^2)F_{P\gamma^*\gamma}(Q_3^2,0),
\end{align}
\noindent
with $Q_3^2 = Q_1^2+Q_2^2+2Q_1Q_2t$. The $I_{1,2}(Q_1,Q_2,t)$ functions are defined in Appendix~\ref{app:a} and $w_{1,2}\equiv w_{1,2}(Q_1,Q_2,t)$. $F_{1,2}$ are products of the relevant TFFs appearing in the blobs in Fig.~\ref{fig:pspole} and are to be evaluated in the SL region. Notice the product of a single- and a double-virtual TFF. The two integrands in Eq.~\eqref{eq:hlblpex} for the $\pi^0$ case are shown in Fig.~\ref{fig:kernel} for a constant TFF ($F_{P\gamma^*\gamma^*}(Q_1^2,Q_2^2)\equiv1$) and $t=0$ along with the $\eta'$ case for the first integrand alone --- similar results hold for different $t$ values\footnote{For a thorough description of these integrands, the interested reader is referred to Ref.~\cite{Nyffeler:2016gnb}.}.

In general, for a given pseudoscalar, the integrand involving $w_2$ is an order of magnitude smaller than that involving $w_1$ (cf. left vs. center panels in Fig.~\ref{fig:kernel}) and is peaked at low energies. Particularly, in such regime it  can be described as  
\begin{equation}
w_2(Q_1,Q_2,t) \sim \frac{Q_{1(2)}^2}{m_P^2}\left( a(t) +b (t)\frac{Q_{2(1)}}{m_{\mu}} + ...\right) + ... \ ,%\mathcal{O}(\frac{Q_1^2}{m^2_{P,\mu}}) ,
\end{equation}
which hints the large rising close to $Q_i^2\sim m_{P,\mu}^2$ (a similar behavior holds for $w_1$), whereas the high-energy behavior reads~\cite{Nyffeler:2016gnb}
\begin{align}
\lim_{Q_{1(2)}^2\to\infty} w_2(Q_1,Q_2,t) &\sim Q_{1(2)}^{-3} + \mathcal{O}(Q_{1(2)}^{-4}), \nonumber \\
\lim_{Q^2\to\infty} w_2(Q,Q,t) &\sim Q^{-4} + \mathcal{O}(Q^{-6}) \label{eq:F2asQ1Q2}.
\end{align}
The integral involving $w_2$ remains therefore finite  even for a constant TFF. As a consequence, only a precise description for the TFFs at very low SL energies is required. By contrast, the dominant integrand, which involves $w_1$, behaves as~\cite{Nyffeler:2016gnb}
\begin{align}
&\lim_{Q_{1}^2\to\infty} w_1(Q_1,Q_2,t) \sim Q_{1}^{-1},\\
&\lim_{Q_{2}^2\to\infty} w_1(Q_1,Q_2,t) \sim Q_{2}^{-2},\\
&\lim_{Q^2\to\infty} w_1(Q,Q,t) \sim Q^{-2}
\end{align}
and constitutes therefore a divergent integral for a constant TFF. Moreover, as it can be observed from Fig.~\ref{fig:kernel} (left and right panels), despite its peak at low energies, this integrand is sensitive to the region above $1$~GeV. This is specially important for heavier pseudoscalars such as the $\eta'$ as it can be observed in Fig.~\ref{fig:kernel} right. In these cases, the low-energy peak is less pronounced and the tail is relevant up to energies beyond $2$~GeV\footnote{The $\pi^0$ features a more pronounced peak as compared to the $\eta$ and $\eta'$ and provides the main contribution. This is related to the chiral enhancement $\propto Q^{-2}$ from the pseudoscalar propagator, which is stronger for the $\pi^0$ given its mass.}

Actually, for the TFFs we employ in Section~\ref{sec:results}, the integral on $Q_1$ and $Q_2$ for $\pi^0,\eta$ and $\eta'$ performed up to $Q_1=Q_2=1$~GeV yields only around $90\%,80\%$ and $70\%$ of the total result, with relative contributions above $95\%$ not reached up to $Q_1=Q_2=1.8,2.5,$ and $3$~GeV, respectively. The discussion above implies the following requirements for a precise calculation (e.g. below $10\%$) when describing the TFFs:
\begin{itemize}
	\item An accurate, precise and ideally model-independent method which can be improved upon via including new theoretical constraints and new experimental data.  
	\item The method should implement a full-energy TFF description for the whole SL region (the time-like (TL) region is not involved in Eq.~\eqref{eq:hlblpex}), including well-known low- and high-energy constraints, the former due to integral weights at low-energies, the latter to render the loop integrals finite.
	\item The method should provide a very precise description at energies as low as $1$~GeV and, at least, a precise description for higher energies up to around $2-3$~GeV. 
\end{itemize}

\begin{figure}[b]
\centering
	\includegraphics[width=0.9\linewidth]{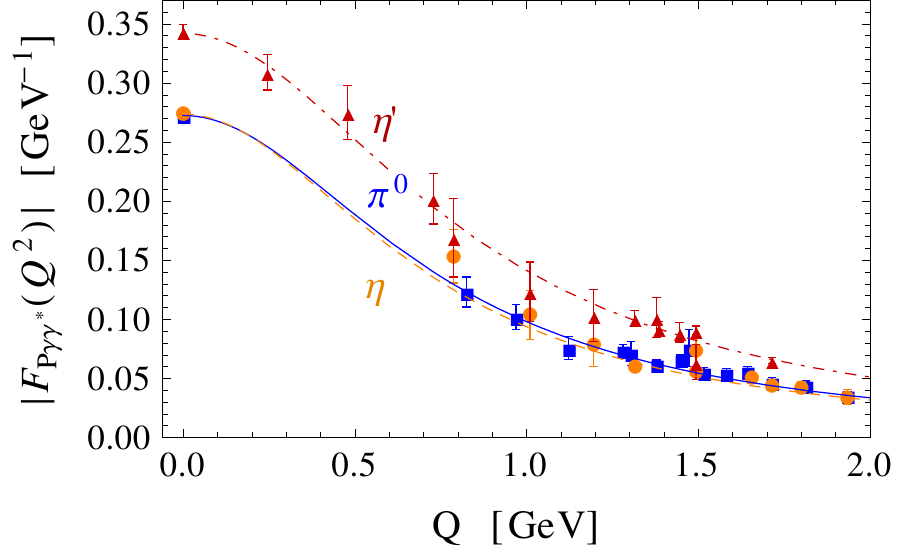}
	\caption{The available low-energy SL data~\cite{Agashe:2014kda,Behrend:1990sr,Gronberg:1997fj,Acciarri:1997yx} for the $\pi^0$ (blue squares), $\eta$ (orange circles) and $\eta'$ (red triangles) TFFs together with our description from Section~\ref{sec:results} as blue, dashed-orange, and dot-dashed-red lines, respectively.}
	\label{fig:data}
\end{figure}

\begin{figure}[b]
\centering
	\includegraphics[width=0.9\linewidth]{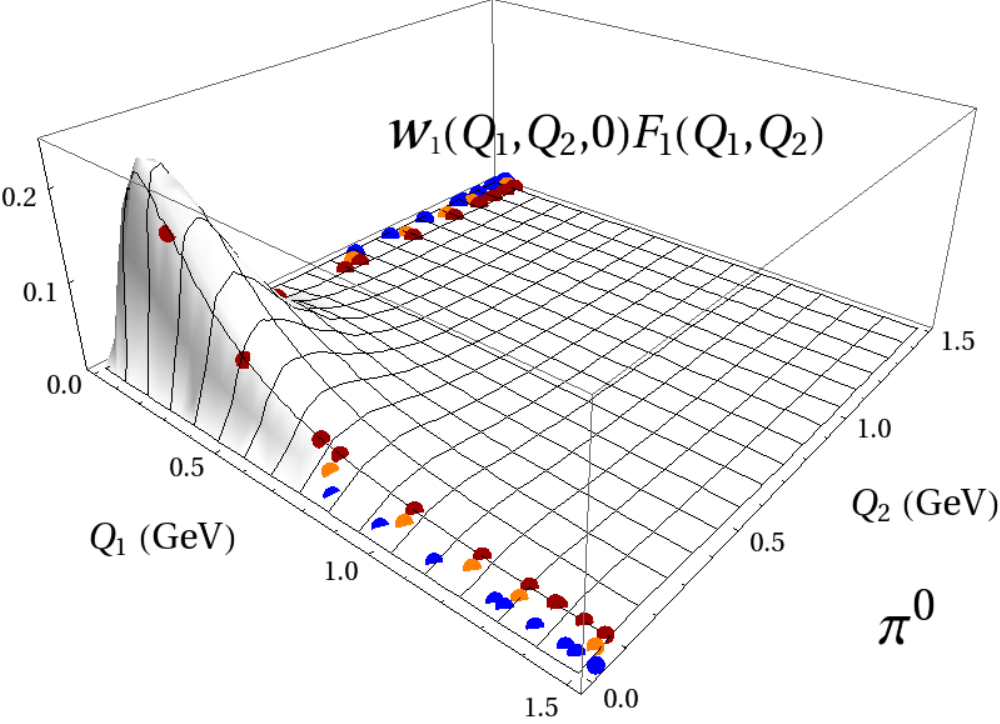}
	\caption{Plot representation of the product of the functions $w_1(Q_1,Q_2,0)F_1(Q_1,Q_2)$ (see Eq.~\eqref{eq:hlblpex}) for $P=\pi^0$. Colored points indicate the $Q$ values for $P=\pi^0,\eta,\eta'$ for which experimental data exist, cf. Fig.~\ref{fig:data}. For ease of illustration, $Q\equiv Q_{1(2)}$ is slightly extrapolated up to the $Q_{2(1)}\neq0$ region. The most important region lacks of experimental data, which manifests  the relevance of the low-energy extrapolation. \label{fig:data2}}
\end{figure}

We believe that none of the current approaches for describing the TFFs fulfill all the criteria enumerated above and an alternative approach is desirable if our goal is a $10\%$ error. The pioneering works in Refs.~\cite{Bijnens:1995xf,Bijnens:2001cq,Hayakawa:1995ps}, which were based on large-$N_c$ or vector meson dominance (VMD) approaches~\cite{Bijnens:1995xf,Bijnens:2001cq,Hayakawa:1995ps}, consisted on a model of the large-$N_c$ limit of QCD. As such, a typical large-$N_c$ error estimate was given to be $30\%$, which represented an adequate error, but it is not enough at the present requested precision. Their systematic uncertainties and achieved accuracy with respect to the real TFFs are difficult to ascribe or systematically improve. A possible venue to refine these approaches is to use them as fitting functions using the current large amount of existing SL data for the single-virtual TFFs (see for instance Refs.~\cite{Knecht:2001qf,Kampf:2011ty,Roig:2014uja} where such approach was pursued), which \textit{could} endow them with certain model independency, or at least, an accuracy beyond the conventional $30\%$ estimate --- a proof of concept is given in Appendix~\ref{app:b}. It is uncertain however up to which accuracy could these approaches describe the real TFFs since the models used to fit are valid only in the large-$N_c$ limit of QCD and, if precision requires, how to systematically improve them. % --- a point to be adressed later within the framework of CAs.
Besides, and unfortunately, there is at present a lack of SL data below $0.8$~GeV \footnote{The exception is the L3 data for the $\eta'$~\cite{Acciarri:1997yx} which, to our best knowledge, has never been used so far in $a_{\mu}^{\textrm{HLbL}}$ calculations.}, see Fig.~\ref{fig:data} and Fig.~\ref{fig:data2}. 
As a result, their low-energy description does not rely on data fitting, but on a fit extrapolation. The precision that such extrapolation provides on the relevant low energies --- even if they may provide an excellent description for the available SL data --- is difficult to quantify. A possible estimate of the eventual precision reached can be obtained by comparing with the available low-energy TL data for the $\eta$ meson~\cite{Arnaldi:2009aa,Berghauser:2011zz,Aguar-Bartolome:2013vpw,Arnaldi:2016pzu,Adlarson:2016hpp}; the accuracy achieved there should provide a reasonable estimate for their SL counterpart. The study performed in Ref.~\cite{Escribano:2015nra} suggests the presence of a non-negligible error for such extrapolations. The lack of ability of these approaches to precisely reproduce the single-virtual low-energy TL data could in addition suggest a similar or even larger uncertainty in the double-virtual region, where no data is available so far to constrain their reconstruction. 

Summarizing, the present data suggest that the standard procedures and the reference studies are not  optimal at low energies and, as we will justify in Sect.~\ref{sec:ca}, they cannot be considered model independent and cannot be systematically improved upon {\emph{up to an arbitrary precision}}. 

More recently, a dispersive reconstruction for the pseudoscalar TFFs has been formulated in Refs.~\cite{Hoferichter:2014vra,Hanhart:2013vba,Xiao:2015uva}. Such approach has the advantage of relying on a data-based framework. As a result, it could in principle be as precise as its required inputs are and its precision systematically improved accordingly. This apparently solves the weakness of previous approaches. However, its full-energy implementation is in practice involved and cannot be complete in a model-independent way. It is for this reason that it is in practice so far limited to the low-energy region up to around $1$~GeV~\cite{Hoferichter:2014vra} and does not incorporate the high-energy constraints in its present form --- the difficulty only increases for the double-virtual description --- the reasons for which we believe them insufficient to pin down the error on the SL integrals beyond $10\%$. Any improvement with respect to traditional approaches comes consequently at the cost of the necessary mid-, and high-energy descriptions. 
It would be desirable then an alternative approach able to deal with all experimental data and theoretical constraints in the SL region (find discussions in Appendix~\ref{app:dr}). 

It was proposed in Ref.~\cite{Masjuan:2007ay} that VMD approaches or the minimal hadronic approximation could be understood, in the large-$N_c$ limit of QCD, within the mathematical framework of Pad\'e approximants (PAs) for the case of meromorphic functions. Such framework would provide the desired systematic method to reconstruct the TFF up to arbitrary precision, improving in accuracy with respect to former VMD approaches. Not only this, PAs guarantee the appropriate low-energy behavior by construction, while allowing to implement, at the same time, the high energies. The application of PAs is however not restricted (as in VMD approaches) to the large-$N_c$ limit of QCD, but can be applied to real $(N_c=3)$ QCD quantities --- a textbook example is the HVP~\cite{Masjuan:2009wy,Aubin:2012me} --- provided they meet certain analytic properties which, for instance, the dispersive approach in Ref.~\cite{Hanhart:2013vba} fulfills. As a consequence, they provide a complementary tool to dispersive approaches in the SL region for improving upon standard VMD descriptions at low energies, but with the advantage of retaining the appropriate mid- and high-energy behaviors as well. Finally, as a difference with respect to all the previous approaches, they provide a systematic error allowing for the desired model independency.

The PAs implementation for the single-virtual pseudoscalar TFFs was discussed for the first time, and in a data-driven way, in Refs.~\cite{Masjuan:2012wy,Escribano:2013kba,Escribano:2015nra,Escribano:2015yup}. More recently, the excellent accuracy at low energies was proved when comparing to the recently released low-energy TL data for the $\eta$ and $\eta'$ mesons~\cite{Escribano:2015nra,Escribano:2015yup}, which corroborated the appropriate description at low energies. This implies in addition that, being based on analytic properties (the latter even allows to reproduce some results in which discontinuities are involved~\cite{Masjuan:2015cjl}), our description  would provide the appropriate extrapolation to the low-energy double-virtual region. However, describing the more general double-virtual TFF requires the extension of Pad\'e approximants to the multivariate case and involves the  use of Canterbury approximants (CAs), which main concepts are illustrated in the following section. This mathematical framework will be the basis for reconstructing the double-virtual TFFs required to calculate the $a_{\mu}^{\textrm{HLbL;}P}$ in a data-driven way in Sect.~\ref{sec:results}.

\section{Canterbury approximants\label{sec:ca}}

\subsection{Definitions}

Given an analytic function, symmetric in its variables, $f(x,y)=f(y,x)$, and with a known formal series expansion, 
\begin{equation}
\label{eq:genseries}
 f(x,y) = \sum_{i,j}c_{i,j}x^iy^j \quad (c_{i,j}=c_{j,i}),
\end{equation}
Canterbury Approximants (CAs)~\cite{Chisholm,Chisholm2,Jones:1976} are defined as rational functions of polynomials $R_N(x,y)$ and $Q_M(x,y)$
\begin{equation}
    C^N_M(x,y) =  \frac{R_N(x,y)}{Q_M(x,y)} =  \frac{\sum_{i,j=0}^{N}  a_{i,j}x^{i}y^{j}}{\sum_{i,j=0}^{M}  b_{i,j}x^{i}y^{j}}  
\end{equation} 
($b_{0,0}=1$ can be generally taken), which coefficients ($i\geq j$) $a_{i,j}\in\mathcal{N}$, and $b_{i,j}\in\mathcal{D}$ are defined as to satisfy the accuracy-through-order conditions~\cite{Masjuan:2015cjl,Masjuan:2015lca}, i.e.,
\begin{align}
    &\sum_{i=0}^{\alpha}\sum_{j=0}^{\beta} b_{i,j}c_{\alpha-i,\beta-j} = a_{\alpha,\beta} \quad \textrm{for } (\alpha,\beta) \in \mathcal{N} , \label{eq:ato} \\
    &\sum_{i=0}^{\textrm{min}(\alpha,M)}\sum_{j=0}^{\textrm{min}(\beta,M)} b_{i,j}c_{\alpha-i,\beta-j} = a_{\alpha,\beta} \quad \textrm{for } (\alpha,\beta) \in \mathcal{E} \notin \mathcal{N}, \nonumber 
\end{align}
where $\textrm{dim}(\mathcal{E}) = \textrm{dim}(\mathcal{N}) + \textrm{dim}(\mathcal{D})$. This is, they are required to match certain terms of the (low-energy) series expansion in Eq.~\eqref{eq:genseries}, which guarantees the correct and desired behavior at low energies. Besides, the high-energy constraints can be included by invoking a two-point CA, which is, as it is standard~\cite{BakerMorris}, requiring a set of accuracy-through-order conditions with the $f(x,y)$ expansion for $x,y\to\infty$\footnote{For a more detailed discussion and examples on how to reconstruct CAs and their performance, we refer the interested reader to the Appendix of Ref.~\cite{Masjuan:2015cjl} --- an additional exercise on convergence is provided in this section below.}. The CAs approach guarantees, among others, the convergence in the cut complex plane of the $C^N_{N+1}(x,y)$ and $C^N_N(x,y)$ as $N\to\infty$\footnote{This means, everywhere except for the TL region above threshold production. Fortunately, this is irrelevant for $g_\mu-2$. Nevertheless, find comments when the TL region is involved in Refs.~\cite{Escribano:2015nra,Escribano:2015yup,Masjuan:2015cjl}.} sequences to meromorphic~\cite{Cuyt:1990} and Stieltjes functions~\cite{Alabiso:1974vk} (see Appendix~\ref{app:stieltjes} for definitions) , which are not only justified in the large-$N_c$ limit of QCD, but also in the light of the dispersive approach from Ref.~\cite{Hanhart:2013vba}, respectively. Note however that the reconstruction outlined above forbids to identify the CA's poles to the physical ones, and rules out typical VMD constructions as a systematic description for the pseudoscalar TFFs, even in the SL region, as previously anticipated\footnote{For a reconstruction of this kind when the function is known to be meromorphic, we refer the interested reader to Ref.~\cite{Masjuan:2007ay}.}.

\subsection{Toy Models}

In order to show the expected performance of CAs, and to provide stronger confidence in our approach, we illustrate its operation, prior to the real case discussion, with the aid of two well-motivated but analytically very different models for the $\pi^0$ TFF\footnote{Our interest in these models resides in the fact that they cannot be well described with a finite set of resonances and the power of PAs is highlighted. The complexity of QCD suggests as well these models to be realistic enough to capture the main QCD ingredients. On top, the convergence of the PA sequence is slow enough to be appreciated in numerical examples.}. The first of them is a large-$N_c$ Regge model~\cite{RuizArriola:2006jge,Arriola:2010aq}, and the second one is an extension of the logarithmic model from Ref.~\cite{Radyushkin:2009zg} to the double-virtual case (see Appendix~\ref{app:b} for a detailed description). For both models convergence is expected since they belong to the class of meromorphic and Stieltjes functions, respectively; it is equally interesting however to test on the convergence rate, a property as relevant for us as convergence itself. 

In the following, we choose the $C^N_{N+1}(Q_1^2,Q_2^2)$ sequence for evaluating $a_{\mu}^{\textrm{HLbL};\pi^0}$; the latter is not only motivated because of convergence properties, but also due to the loop integral in Eq.~\eqref{eq:hlblpex}, which requires a $C^N_M(Q_1^2,Q_2^2)$ sequence for which $M>N$ if integration is to be taken up to infinity. 

To illustrate the different possibilities that CAs offer, we proceed as follows. First of all, we reconstruct the TFFs models employing only the information which is provided from the low-energy TFF expansion, Eq.~\eqref{eq:genseries}, alone. This reconstruction does not incorporate however the high-energy constraints which can be expressed as\footnote{For the physical TFF, it is possible to implement the single-virtual high-energy asymptotic or Brodsky-Lepage (BL) behavior as well, see Sect.~\ref{sec:results}. Unfortunately, this is not possible for the present models, which have a logarithmic enhancement --- see Appendix~\ref{app:b}. In any case, we note that the $C^N_{N+1}(Q_1^2,Q_2^2)$ approximant already implements the correct $Q^{-2}$ BL behavior.} 
\begin{equation}
 F_{P\gamma^*\gamma^*}^{\textrm{Regge(Log)}}(Q^2,Q^2) = \mathcal{C}_1Q^{-2} + \mathcal{C}_2Q^{-4} + ... \ . \label{eq:HEexp}
\end{equation}

Indeed, the chosen sequence behaves for large energies as $C^N_{N+1}(Q^2,Q^2)\sim Q^{-4}$ rather than $Q^{-2}$. Consequently, we implement in a second step the conditions above in a sequential manner, starting with the $Q^{-2}$ high-energy behavior (but not constraining the particular $\mathcal{C}_1$ value above) and progressively including additional $\mathcal{C}_i$ coefficients.

The results for the approximants, which are shown in Table~\ref{tab:model} along with the models' exact result, illustrate the following features
\begin{itemize}
\item The first approach (LE row of Table~\ref{tab:model}) shows a clear convergence according to our expectations; indeed, the second $N=1$ element already provides a small systematic error, well below $10\%$, for both models. 
\item Implementing the $Q^{-2}$ high-energy behavior (OPE$_0$ row) shows a clear improvement on convergence, which is expected given the high-energy behavior of the integrand in Eq.~\eqref{eq:hlblpex}.
\item Further constraining the $\mathcal{C}_1$ and $\mathcal{C}_2$ coefficients (OPE$_1$ and OPE$_2$ rows) does not seem to affect or change convergence, which points to the relevance of the low energies and to the fast convergence of the method\footnote{For the logarithmic model, $F_{P\gamma^*\gamma^*}^{\textrm{log}}(Q^2,Q^2)=F_{P\gamma\gamma}\frac{M^2}{M^2+Q^2}$. Consequently, all the OPE coefficients, $\mathcal{C}_n$, are trivially satisfied within our approach.}. 
\item Besides the particular systematic error for each model, we observe an expected more general feature: the systematic error of a given element $N$ can be inferred from its difference with respect to the $N-1$ element. This provides a model-independent estimation for the systematic uncertainty and, thereby, the sought model-independent result.
\end{itemize}
\begin{table}[t]
\centering
\begin{tabular}{ccccc@{\hskip0.5cm}cccc} \toprule
 & \multicolumn{4}{c}{Regge Model}  & \multicolumn{4}{c}{Log Model} \\
  & $C^0_1$ & $C^1_2$ & $C^2_3$ & $C^3_4$ & $C^0_1$ & $C^1_2$ & $C^2_3$ & $C^3_4$ \\ \hline
LE & $55.2$ & $59.7$ & $60.4$ & $60.6$ & $56.7$ & $64.4$ & $66.1$ & $66.8$ \\
OPE$_0$ & $65.7$ & $60.8$ & $60.7$ & $60.7$ & $65.7$ & $67.3$ & $67.5$  & $67.6$  \\  
OPE$_1$ & $-$ & $60.6$ & $60.7$ & $60.7$ & $65.7$ & $67.3$ & $67.5$  & $67.6$  \\  
OPE$_2$ & $-$ & $60.8$ & $60.7$ & $60.7$ & $65.7$ & $67.3$ & $67.5$  & $67.6$  \\  \hline
Fact & $54.6$ & $57.3$ & $57.4$ & $57.5$ & $54.6$ & $60.3$ & $61.3$ & $61.6$ \\
Fit$^{\textrm{OPE}}$ & $66.3$ & $62.7$ & $61.1$ & $60.8$ & $79.6$ & $71.9$ & $69.3$ & $68.4$ \\ \hline
Exact  & \multicolumn{4}{c}{$60.7$}  & \multicolumn{4}{c}{$67.6$} \\ \botrule
\end{tabular}
\caption{The results for $a_{\mu}^{\textrm{HLbL};\pi^0}\times10^{11}$ using the Regge and logarithmic models (last row) are compared to their $C^N_{N+1}(Q_1^2,Q_2^2)$ sequence of approximants' results.  The LE row uses a pure low-energy reconstruction, whereas the OPE$_n$ rows incorporate high-energy constraints. The Fact row serves as an illustration of what a factorization approach would have yield. Finally, Fit$^{\textrm{OPE}}$ row shows what a $C^N_{N+1}(Q_1^2,Q_2^2)$-like fitting function with the appropriate OPE behavior would lead. More details in the main text.\label{tab:model}}
\end{table}
In addition, it is worth to comment on factorization approaches for which $C^N_{N+1}(Q_1^2,Q_2^2) \sim C^N_{N+1}(Q_1^2,0)\times C^N_{N+1}(Q_2^2,0)$. These are very popular and seem to represent a good approximation at low energies~\cite{Maris:2002mz,Bijnens:2012hf,Masjuan:2015lca,Xiao:2015uva} (note that non-factorizable effects are formally of order $(Q^2)^2$ in the low-energy expansion in any case). The results are shown in the sixth row of Table~\ref{tab:model} (Fact row) and show a potential large systematic error. The latter is however not only due to the wrong behavior at high-energies --- our low-energy reconstruction in the second row of Table~\ref{tab:model} (LE row) does not fulfill it either ---  but to the fact that not even the series expansion factorizes.

Finally, in our discussion above, it cannot be overemphasized the relevance of having employed the low-energy expansion Eq.~\eqref{eq:genseries} when reconstructing the approximants --- as the framework requires --- rather than fitting the rational functions to data themselves. To illustrate this statement, we show in the last row of Table~\ref{tab:model} (Fit$^{\textrm{OPE}}$) what would have been obtained if fitting the $C^N_{N+1}(Q_1^2,Q_2^2)$ rational functions, with the OPE behavior implemented, to a double-virtual data grid ranging from $0\leq Q_{1,2}^2 \leq 35~\textrm{GeV}^2$ with a $2.3~\textrm{GeV}^2$ spacing. The obtained convergence is slower, and illustrates the difference and the power of CAs with respect to standard fitting approaches. 

Summarizing the previous results: we find that the $C^N_{N+1}(Q_1^2,Q_2^2)$ sequence of approximants provides an excellent convergence when calculating $a_{\mu}^{\textrm{HLbL};\pi^0}$ for the chosen TFF models --- which is further accelerated if the high-energy behavior is accounted for. 
More important, we find that the systematic uncertainty can be estimated from the difference among the elements within the sequence, which represents the main advantage from our approach and provides for the model independency.
Having introduced CAs, motivated a sequence  and illustrated its performance, we proceed to apply this approach for the real QCD case.

\section{Results\label{sec:results}}

For the physical TFF, we define the formal series expansion, Eq.~\eqref{eq:genseries}, in terms of the low-energy parameters (LEPs) $b_P,c_P, a_{P;1,1},...$ as
\begin{multline}
\label{eq:TFFseries}
F_{P\gamma^*\gamma^*}(Q_1^2,Q_2^2) =F_{P\gamma\gamma}(0,0)\left(  1-\frac{b_P}{m_P^2}(Q_1^2+Q_2^2)  \right. \\ 
\left. + \frac{c_P}{m_P^4}(Q_1^4+Q_2^4) + \frac{a_{P;1,1}}{m_P^4}Q_1^2Q_2^2 + ... \right).
\end{multline}

It turns out that, under certain approximations, the authors of Ref.~\cite{Stollenwerk:2011zz} proved the isovector contribution to the TFF to be a Stieltjes function (cf. Appendix~\ref{app:stieltjes}), for which convergence of Pad\'e approximants is guaranteed in advance.

Actually, Pad\'e theory not only provides a convergence theorem for a sequence of PAs to Stieltjes functions, i.e.,  $\lim_{N,M\to \infty} P^N_M(s) - f(s) =0$, but also its rate of convergence~\cite{BakerMorris,Masjuan:2008cp,Masjuan:2009wy}, which is given by the difference of two consecutive elements in the PA sequence~\cite{Masjuan:2008fv,Masjuan:2012wy,Escribano:2013kba,Escribano:2015nra}.

Furthermore, in the large-$N_c$ limit of QCD, the TFF becomes a meromorphic function, for which convergence is guaranteed as well~\cite{Masjuan:2007ay,Masjuan:2008fr}. The sum rule approach employed in Ref.~\cite{Klopot:2012sa} for describing the TFF is again of the Stieltjes kind. Moreover, our experience from analyses of the TFF from the SL data~\cite{Masjuan:2012wy,Escribano:2013kba} and the excellent predictions achieved in the low-energy TL region~\cite{Escribano:2015nra,Escribano:2015yup} suggests that convergence to the TFF is at work and that its relevant analytical properties are retained. We understand that all these features hold for the double-virtual case too.

The available analytical information on the TFF is scarce though; at low energies $F_{P\gamma\gamma}(0,0)$ is theoretically related in the chiral (and large-$N_c$ for the $\eta$ and $\eta'$) limit to the Adler~\cite{Adler:1969gk}-Bell-Jackiw~\cite{Bell:1969ts} anomaly, and can be expressed as\footnote{$\mathcal{Q}$ stands for the charge matrix, $F$ is the pion decay constant in the chiral limit, and $\lambda_{\pi,\eta,\eta'}=\lambda^{3,8,0}$ in the chiral limit with $\lambda^a$ the Gell-Mann matrices and $\lambda^0=\sqrt{2/3}~\mathds{1}_{3\times3}$.}
\begin{equation}
 F_{P\gamma\gamma}(0,0) = \frac{N_c}{4\pi^2F}\operatorname{tr}(\mathcal{Q}^2\lambda_P).
\end{equation}

This expression is, strictly speaking, valid only at the LO in both the chiral and the large-$N_c$ limits of QCD. Corrections to it which 
involve, among others, the $\eta-\eta'$ mixing at the given order~\cite{Bickert:2016fgy} are calculated in terms of unknown low-energy constants~\cite{Goity:2002nn}. For this reason, we use instead the experimental results for $P\to\gamma\gamma$ decays in order to avoid model dependencies, which relation to $F_{P\gamma\gamma}(0,0)$ follows from $|F_{P\gamma\gamma}(0,0)| = \sqrt{\frac{64\pi}{(4\pi\alpha)^2}\frac{\Gamma(P\to\gamma\gamma)}{m_P^3}}$.

At small but finite virtualities, there are no further available theoretical predictions, and higher LEPs in Eq.~\eqref{eq:TFFseries} are theoretically unknown. Still, some of these LEPs were extracted for the single-virtual case in a data-driven approach using PAs~\cite{Masjuan:2012wy,Escribano:2013kba,Escribano:2015nra,Escribano:2015yup} which, as said, have proven extremely accurate when confronting them against the low-energy TL data. A similar procedure for the most general double-virtual case would be possible once double-virtual experimental data becomes available. At  high energies, the TFF behavior can be theoretically described within pQCD. For the case of a single-virtual photon the TFF is known to behave according to the Brodsky-Lepage (BL)~\cite{Lepage:1980fj} asymptotic behavior
\begin{equation}
\label{eq:bl}
\lim_{Q^2\to\infty} F_{P\gamma^*\gamma}(Q^2,0) = P_{\infty}Q^{-2} + \mathcal{O}(Q^{-4}),
\end{equation}
where $\pi_{\infty}$ depends on the pion decay constant, whereas $\eta_{\infty}$ and $\eta'_{\infty}$ depend on the mixing parameters and the singlet-axial current running effects~\cite{Escribano:2015nra,Escribano:2015yup}. For the double-virtual case, the TFF behavior at high energies is obtained from the operator product expansion (OPE)~\cite{Novikov:1983jt,Jegerlehner:2009ry}
\begin{equation}
\label{eq:ope}
\lim_{Q^2\to\infty} F_{P\gamma^*\gamma^*}(Q^2,Q^2) = \frac{P_{\infty}}{3}\left( \frac{1}{Q^2} -\frac{8}{9}\frac{\delta_P^2}{Q^4} \right)+ \mathcal{O}(Q^{-6}),
\end{equation}
where the numerical values for the parameters introduced in Eqs.~(\ref{eq:TFFseries}, \ref{eq:bl}, \ref{eq:ope}) can be found in Table~\ref{tab:input}. Remarkably, Eqs.~\eqref{eq:bl} and~\eqref{eq:ope} ensure the convergence of the integrands in Eq.~\eqref{eq:hlblpex} and suggest the use of the  $C^N_{N+1}(Q_1^2,Q_2^2)$ sequence explored in the previous section. Given the current available information on the double-virtual TFF, only the first two elements can be reconstructed. They are expressed as 
\begin{widetext}
\begin{align}
 C^0_1(Q_1^2,Q_2^2) &= \frac{F_{P\gamma\gamma}(0,0)}{1+\frac{b_P}{m_P^2}(Q_1^2+Q_2^2)}, \label{eq:c01}\\
 C^1_2(Q_1^2,Q_2^2) &= \frac{F_{P\gamma\gamma}(0,0)(1+\alpha_{1}(Q_1^2+Q_2^2) + \alpha_{1,1}Q_1^2Q_2^2)}{1+\beta_{1}(Q_1^2+Q_2^2)+\beta_{2}(Q_1^2+Q_2^2)+\beta_{1,1}Q_1^2Q_2^2+\beta_{2,1}Q_1^2Q_2^2(Q_1^2+Q_2^2)}.\label{eq:c12}
\end{align}
\end{widetext}

The connection to the LEPs from Eq.~\eqref{eq:TFFseries} is already visible in Eq.~\eqref{eq:c01}; the relation of the $\alpha_{i,j}$ and $\beta_{k,l}$ parameters in Eq.~\eqref{eq:c12} to the LEPs is involved enough as not to fit in a single line. First, the single-virtual parameters $F_{P\gamma\gamma}(0,0),\alpha_1,\beta_1$ and $\beta_2$ must be reconstructed. $F_{P\gamma\gamma}(0,0)$ is related, as mentioned,  to the $P\to\gamma\gamma$ decays and can be extracted from the experimental values in Ref.~\cite{Agashe:2014kda}; $\alpha_1,\beta_1$ and $\beta_2$ are related to the linear, quadratic and cubic terms in the single-virtual low-energy expansion, Eq.~\eqref{eq:TFFseries}. These three parameters have been extracted from a data-driven approach in Refs.~\cite{Escribano:2013kba,Escribano:2015nra,Escribano:2015yup} for the $\eta$ and $\eta'$, where they have been referred to as $b_Pm_P^{-2}, c_Pm_P^{-4}$ and $d_Pm_P^{-6}$, respectively. Alternatively, the cubic term can be traded for the BL asymptotic behavior, which is extremely convenient for the $\pi^0$ given the precise theoretical prediction (which contrasts with the $\eta$ and $\eta'$ cases, see Table~\ref{tab:input}). Consequently, for the $\pi^0$, we employ the linear and quadratic terms determined in Ref.~\cite{Masjuan:2012wy}\footnote{In the near future, the data which are being analyzed at BES III~\cite{Adlarson:2014hka} 
Collaboration in the low-energy SL region will allow for an accurate extraction of $d_{\pi}$.} together with the BL prediction, which implies $\lim_{Q^2\to\infty}Q^2C^1_2(Q^2,0)=2F_{\pi}=0.1884(3)$~GeV.
It remains to determine the double-virtual parameters $\alpha_{1,1},\beta_{1,1}$ and $\beta_{1,2}$. For the $\pi^0$ case, it is possible to relate two of them to the high-energy expansion $P_{\infty}$ and $\delta_{P}$ parameters in Eq.~\eqref{eq:ope}. For the $\eta$ and $\eta'$, $\delta_P$ is unknown; we take $\delta_{\eta,\eta'}=\delta_{\pi}$ and ascribe an extra  $30\%$ systematic error from $SU(3)_F$-breaking (and large-$N_c$) effects\footnote{We note that such error covers for the observed $\pi^0,\eta$ and $\eta'$ differences for all the parameters which have been determined so far: $F_{P\gamma\gamma}(0,0), b_Pm_P^{-2}, P_{\infty}, ...$~. Besides, we find that, in practice, the $a_{\mu}^{\textrm{HLbL};P}$ dependence on this parameter is certainly mild.}. Finally, one parameter remains to be determined. This can and should be related to the low-energy parameter $a_{P;1,1}$ in Eq.~\eqref{eq:TFFseries}, which could be determined if double-virtual data becomes available. Hopefully, this may be possible in the future at BES III~\cite{Adlarson:2014hka} for the $\pi^0$. Additional sources of information would be $P\to\bar{\ell}\ell$~\cite{Masjuan:2015lca,Masjuan:2015cjl} and $P\to\bar{\ell}\ell\bar{\ell'}\ell'$ decays~\cite{Escribano:2015vjz}. Unfortunately, the available experimental precision for these measured decays is still not enough to make an extraction --- find more comments later Section~\ref{sec:data}. For this reason, and to be as model independent as possible, we take for $a_{P;1,1}$ the most general range which is physically accessible without spoiling the TFF's analytic properties, i.e., avoiding the presence of poles for the $C^1_2(Q_1^2,Q^2_2)$ in the SL region. This leads in practice to a range of the kind $a_{P;1,1}^{\textrm{min}} \leq a_{P;1,1} \leq a_{P;1,1}^{\textrm{max}}$\footnote{Particularly, we find that $a_{\pi;1,1}^{\textrm{min}}=1.89b_{\pi}^2$, $a_{\eta;1,1}^{\textrm{min}}=1.65b_{\eta}^2$, and $a_{\eta';1,1}^{\textrm{min}}=1.32b_{\eta'}^2$, whereas $a_{\pi;1,1}^{\textrm{max}}=2.10b_{\pi}^2$, $a_{\eta;1,1}^{\textrm{max}}=6.00b_{\eta}^2$ and $a_{\eta';1,1}^{\textrm{max}}=3.41b_{\eta'}^2$.} and completes the discussion about the CAs reconstruction.

The numerical integrals have been calculated with {\textit{Mathematica 8.0} using the {\texttt{AdaptiveQuasiMonteCarlo}} method. The obtained results for the $C^0_1(Q_1^2,Q_2^2)$ are collected, for the different pseudoscalars, in Table~\ref{tab:gm2} first column; the results for the $C^1_2(Q_1^2,Q_2^2)$ are collected, for the $a_{P;1,1}$ considered range, in the second and third columns of Table~\ref{tab:gm2}. 
\begin{table}[t]
\begin{tabular}{cccc} \toprule
	$a_{\mu}^{\textrm{HLbL;}P}$  & $C^0_1$  & $C^1_2[a_{P;1,1}^{\textrm{min}}]$ & $C^1_2[a_{P;1,1}^{\textrm{max}}]$ \\ \hline
	$\pi^0$  & $65.3(1.4)(2.4)[2.8]$  &  $64.1(1.3)[1.3]$ & $63.0(1.1)(0.5)[1.2]$ \\ 
	$\eta$  & $17.1(0.6)(0.2)[0.6]$  &  $16.3(0.8)[0.8]$ & $16.2(0.8)(0.6)[1.0]$ \\ 
	$\eta'$  & $16.0(0.5)(0.3)[0.6]$  & $14.7(0.7)[0.7]$ & $14.3(0.5)(0.5)[0.7]$ \\ \hline
	Total  &  $98.4[2.9]$  &  $95.1[1.7]$ & $93.5[1.7]$ \\ \botrule
\end{tabular}
\caption{Our $a_{\mu}^{\textrm{HLbL};P}$ results in units of $10^{-11}$. The second column is our simplest $C^0_1(Q_1^2,Q_2^2)$ approximant; the third and fourth columns refer to the $C^1_2(Q_1^2,Q_2^2)$ one, and stand for the lower and upper $a_{P;1,1}$ values respectively. See description in the text.  \label{tab:gm2}}
\end{table}  
The errors are statistical only and arise from a MC analysis. For the $C^0_1$ case, the first and second errors arise from the TFF normalization ($F_{P\gamma\gamma}(0,0)$) and the slope parameter ($b_P$), respectively; for the $C^1_2$ $a_{P;1,1}^{\textrm{min}}$ choice (second column in Table~\ref{tab:gm2}), the error is due to the single virtual parameters $F_{P\gamma\gamma}(0,0),b_P,c_P,d_P$ and $P_{\infty}$\footnote{In this limiting case, the $\alpha_{1,1}$ and $\beta_{2,1}$ parameters in Eq.~\eqref{eq:c12} vanish, and the $\delta_P$ parameter becomes as a consequence irrelevant.};  for the $C^1_2$ $a_{P;1,1}^{\textrm{max}}$ choice, an additional error related to $\delta_P$ arises; the last error in brackets stands in every case for the combination in quadrature of the individual ones. 

The last line, \textit{Total}, is the sum of the $\pi^0,\eta$ and $\eta'$ contributions. To estimate the systematic uncertainty, we take the largest difference among the $a_{P;1,1}^{\textrm{min}}$ and $a_{P;1,1}^{\textrm{max}}$ choices with respect to the $C^0_1(Q_1^2,Q_2^2)$, which even if it may overestimate the systematic uncertainty, still allows to reach an error below $10\%$. We note that this procedure corresponds to assume a fully correlated systematic error among the different pseudoscalars systematic uncertainties. Our final result is:

\begin{align}
a_{\mu}^{\textrm{HLbL};P} &=  (93.5\div 95.1)(1.7)_{\textrm{stat}}(4.9)_{\textrm{sys}}\times10^{-11} \nonumber \\
	&\to 94.3(5.3) \times10^{-11}, \label{eq:final}
\end{align}
where in the last line, the mean value among the third and fourth column in Table~\ref{tab:gm2} has been taken, and their difference associated as an additional source of uncertainty which has been added in quadrature to the previous ones. In Appendix~\ref{app:beyondpole} we combine our final result with the \textit{Glasgow consensus} in an attempt to provide a value for the whole $a_{\mu}^{\textrm{HLbL}}$.

Our obtained result can be compared to former determinations of the pseudoscalar-pole contribution, which appear in Refs.~\cite{Knecht:2001qf,Roig:2014uja}. The first of them~\cite{Knecht:2001qf}, which was intended to clarify a sign discussion, reads $a_{\mu}^{\textrm{HLbL;}P} = (58(10)+13(1)+12(1)=83(12))\times10^{-11}$ and has a crude error estimation as discussed in Ref.~\cite{Knecht:2001qf}. 
Particularly, the $\eta$ and $\eta'$ contributions were provided as an order of magnitude estimate, and for that reason only a simplified (and factorized) model was used. In addition, we note that our approach incorporates far more data for the TFFs which have become available since this study appeared. 

The second and more recent study, Ref.~\cite{Roig:2014uja}, is based on resonance chiral theory and obtains $a_{\mu}^{\textrm{HLbL;}P} = (57.5(6)+14.4(2.6)+10.8(0.9)=82.7(6.6))\times10^{-11}$. Their analysis makes use of a similar data set to that employed in our approach for the $\pi^0$ case, but only up to 2014. The difference for the $\pi^0$ contribution could be ascribed to their different high-energy behavior since they cannot incorporate both BL and OPE, Eqs.~\eqref{eq:bl},~\eqref{eq:ope} --- a known issue when using a single resonance~\cite{Masjuan:2007ay}. For the $\eta$ and $\eta'$ cases, their approach description becomes more involved, as an appropriate $\eta-\eta'$ mixing description requires an analysis at the next-to-leading order within their approach, with an error difficult to quantify. For this reason, Ref.~\cite{Roig:2014uja} uses instead $U(3)_F$ symmetry arguments to relate both $\eta$ and $\eta'$ TFFs with the $\pi^0$ one. The differences we find illustrate the importance of a data-based approach to describe these TFFs. 

As a final remark, both approaches in Refs.~\cite{Knecht:2001qf,Roig:2014uja}, which rely on the large-$N_c$ limit, can be understood as a CA to meromorphic functions. Then,  the (missing) systematic error for their reconstruction --- in which the poles are fixed in advance --- is larger than in our case, see Ref.~\cite{Masjuan:2007ay} and comments in Appendix~\ref{app:b}. Moreover, they do not employ a low-energy reconstruction, but a fitting procedure which, as illustrated in Sec.~\ref{sec:ca}, entails an even larger error. Overall, these considerations suggest total unaccounted errors above $10\%$ for these approaches. An additional calculation based on the VMD models from Ref.~\cite{Knecht:2001qf} but fitted to lattice simulations for the $\pi^0$ TFF obtained $a_{\mu}^{\textrm{HLbL};\pi}=65.0(8.3)\times10^{-11}$~\cite{Gerardin:2016cqj}. The latter includes statistical error, but lacks the inclusion of a systematic uncertainty inherent to the large-$N_c$-based fitted model.

\section{The role of future data\label{sec:data}}

Finally, we outline the impact that future data would have in our TFFs reconstruction and thereby in our $a_{\mu}^{\textrm{HLbL;}P}$ determination. First of all, the most relevant parameters are the values of the TFFs at zero virtualities. To see this, note from Eq.~\eqref{eq:hlblpex} that the whole contribution is proportional to the square of this quantity. As such, a relative $\Delta_{F_{P\gamma\gamma}}$ error on the former directly translates into a  $2\Delta_{F_{P\gamma\gamma}}$ relative error for $a_{\mu}^{\textrm{HLbL;}P}$. In this respect, and in the light of Table~\ref{tab:gm2}, a further reduction on this quantity will substantially improve our total error. The PrimEx-II experiment at JLab~\cite{Gasparian:2016oyl} and the experiment planned at KLOE-2~\cite{Babusci:2011bg} on the $\gamma\gamma\rightarrow\pi^0\rightarrow\gamma\gamma$ reaction will halve the $\pi^0$ contribution error. Regarding the $\eta$ and $\eta'$, the future {\textit{GlueX}} experiment at JLab~\cite{Dudek:2012vr} would allow to reduce the $\eta$ and $\eta'$ counterpart associated error --- only a $3\%$ final precision for the $\eta$ case has been reported so far~\cite{Gan:2015nyc}, which would again halve the current error.

Second, from Section~\ref{sec:results} and the results in Table~\ref{tab:gm2}, we find pressing to get new data on the $\pi^0$ TFF. This would allow to improve our $b_{\pi}$ and $c_{\pi}$ determinations --- especially the dominant systematic error, see Ref.~\cite{Masjuan:2012wy} --- and to extract the $d_{\pi}$ parameter, providing therefore an alternative single-virtual description in terms of LEPs alone and no high energy coefficients, which would be a valuable cross-check of our results. This will be possible in the near future once the ongoing analysis at BES III~\cite{Adlarson:2014hka} Collaboration becomes published --- in addition, further data is expected at low energies from KLOE-2~\cite{Babusci:2011bg} and {\textit{GlueX}} experiment~\cite{Gan:2015nyc}.

In addition, in view of the discussion raised by \babar~\cite{Aubert:2009mc} data concerning the TFF behavior at high energies, an interesting test could be to reconstruct the $\pi^0$ TFF from the light-quark content of the $\pi^0$ meson. From our knowledge of the $\eta$ and $\eta'$ TFFs and the $\eta-\eta'$ mixing in the flavor basis it is possible to extract from experimental data a pure light-quark TFF and use it to calculate what we denominate the light-quark $a_{\mu}^{\textrm{HLbL;}l.q.}$ which should be similar to the actual $\pi^0$ one. Appendix~\ref{app:lq} contains the detail of such calculation and the results reported there support our final result from Eq.~(\ref{eq:final}).
 
Furthermore, there are efforts to measure the $\eta$ and $\eta'$ TFFs at low SL energies at BES III~\cite{Adlarson:2014hka} --- a similar study would be possible at the {\textit{GlueX}} experiment as well~\cite{Dudek:2012vr,Gan:2015nyc}. 

Finally, it remains an important task to get the first information on the double-virtual TFF given the $a_{P;1,1}$ parameter-induced error. This could be possible for the $\pi^0$ at BES III~\cite{Adlarson:2014hka}. Such a measurement would not only allow to improve on our current estimate, but to eventually obtain further double-virtual parameters. In addition, the latter would allow to trade the OPE expansion parameters in Eq.~\eqref{eq:ope} in favor of the LEPs in Eq.~\eqref{eq:TFFseries} when reconstructing the TFF. As said, it is possible to use as well $P\to\bar{\ell}\ell$ decays. However, the current precision is not sufficient to provide a competitive constraint, see Appendix~\ref{app:pll} for details. Similarly, $P\to\bar{\ell}\ell\bar{\ell'}\ell'$ decays provide an interesting potential source of information. Unfortunately, once more, a high precision is required for them in order to provide competitive constraints --- around a $5\%$ precision for the yet not measured $\eta^{(\prime)}\to2\mu^+2\mu^-$ decays, see Ref.~\cite{Escribano:2015vjz}. Alternatively, there is the possibility that, in the future, lattice simulations at the physical $\pi^0$ mass could provide valuable information in determining the LEPs in our method --- see advances in Refs.~\cite{Feng:2012ck,Cohen:2008ue,Lin:2013im,Gerardin:2016cqj}.

Finally, our approach could incorporate the low-energy parameters predicted from dispersive theory~\cite{Hanhart:2013vba,Hoferichter:2014vra,Xiao:2015uva}. Not only that, but our approach would allow to extend these frameworks at higher energies, providing an analytic continuation to the high-energy region, where these approaches cannot apply (find further comments in Appendix~\ref{app:dr}).

\section{Conclusions}

In this work, we have employed the mathematical framework of Canterbury approximants (bivariate Pad\'e approximants) in order to reconstruct the double-virtual pseudoscalar TFFs and calculate the pseudoscalar-pole contribution to $a_{\mu}^{\textrm{HLbL}}$. The method allows to incorporate, at the same time, both the low- and the high-energy information on the TFFs. Though the former play the most relevant role in this calculation, a precise calculation should consider both simultaneously. This feature represents the first advantage from our method with respect to resonance approaches or dispersive representations. 

The required information for the reconstruction of CAs has been obtained from our works in Refs.~\cite{Masjuan:2012wy,Escribano:2013kba,Escribano:2015nra,Escribano:2015yup} and employs data from over 13 different collaborations. As a novelty of our approach, the method provides a systematic treatment --- which advantage with respect to resonance approaches is especially obvious when dealing with the $\eta$ and $\eta'$ --- and allows for a systematic error estimation, which provides for the model independence of the result and the second advantage with respect to existing approaches.

As a result, we have found $a_{\mu}^{\textrm{HLbL};P} = 94.3(5.3)\times10^{-11}$, which is larger than previous estimates by a quantity which essentially corresponds to future experiments' uncertainty. 
We note that such quantity is extremely interesting, as it is not only the dominating contribution to $a_{\mu}^{\textrm{HLbL}}$, but it is present in any data-driven approach for calculating $a_{\mu}^{\textrm{HLbL}}$ so far. Furthermore, our approach will benefit in the near future from the large amount of data which is expected to appear, including the SL one which, so far, is not included in dispersive approaches. Concerning the latter, our approach can also benefit from the TFF's dispersive representations, as soon as the LEPs are reported, to account for the $\pi^0,\eta$ and $\eta'$ contributions to $a_\mu$, which shows the flexibility and complementarity of our approach

\begin{acknowledgments}

The authors would like to thank A. Nyffeler and K. Kampf for discussions and to S. Leupold for comments concerning the dispersive representation. Work supported by the Czech Science Foundation (grant no. GACR 15-18080S).

\end{acknowledgments}

\appendix

\section{Functions involved in the HLbL}
\label{app:a}

The pseudoscalar-pole contribution to the HLbL tensor is given as~\cite{Knecht:2001qf}
\begin{multline}
\Pi_{\mu\nu\lambda\rho}^{P-\textrm{pole}}(q_1,q_2,q_3) = \\
  i\frac{F_{P\gamma^*\gamma^*}(q_1^2,q_2^2)F_{P\gamma^*\gamma^*}(q_3^2,k^2)}{(q_1+q_2)^2-m_P^2}
               \epsilon_{\mu\nu\alpha\beta}q_1^{\alpha}q_2^{\beta}\epsilon_{\lambda\rho\sigma\tau}q_3^{\sigma}k^{\tau}  \\
 +  i\frac{F_{P\gamma^*\gamma^*}(q_1^2,k^2)F_{P\gamma^*\gamma^*}(q_3^2,q_2^2)}{(q_2+q_3)^2-m_P^2} 
            \epsilon_{\mu\rho\alpha\beta}q_1^{\alpha}k^{\beta}\epsilon_{\nu\lambda\sigma\tau}q_2^{\sigma}q_3^{\tau}  \\
 +  i\frac{F_{P\gamma^*\gamma^*}(q_1^2,q_3^2)F_{P\gamma^*\gamma^*}(k^2,q_2^2)}{(q_1+q_3)^2-m_P^2} 
            \epsilon_{\mu\lambda\alpha\beta}q_1^{\alpha}q_3^{\beta}\epsilon_{\nu\rho\sigma\tau}q_2^{\sigma}k^{\tau},  \label{eq:polehlbl}
\end{multline}
where $\epsilon^{0123}=+1$, $q_1,q_2,q_3$ are outgoing from the blob depicted in Fig.~\ref{fig:HLbL} and $k=q_1+q_2+q_3$ is incoming to it.

The $I_{1,2}(Q_1,Q_2,t)$ functions involved in the $a_{\mu}^{\textrm{HLbL};P}$ calculation, Eq.~\eqref{eq:hlblpex}, are defined as
\begin{multline}
I_1(Q_1,Q_2,t) =  \frac{-1}{m_{\mu}^2Q_3^2}\Bigg[ \frac{4m_{\mu}^2t}{Q_1Q_2} \\
+(1-R_{m_1})\bigg( \frac{2Q_1t}{Q_2} \-4(1-t^2) \bigg)  - (1-R_{m_1})^2\frac{Q_1t}{Q_2} \\ - 8X(Q_1,Q_2,t)( Q_2^2 -2m_{\mu}^2)(1-t^2) \Bigg], 
\end{multline}
\begin{multline}
I_2(Q_1,Q_2,t) =  \frac{-1}{m_{\mu}^2Q_3^2}\Bigg[ 2(1-R_{m_1})\bigg(\frac{Q_1t}{Q_2}+1\bigg)  \\ +2(1-R_{m_2})\bigg(\frac{Q_2t}{Q_1}+1\bigg) \\ + 4X(Q_1,Q_2,t)\left(Q_3^2 +2m_{\mu}^2(1-t^2) \right)\Bigg],
\end{multline}
where the following functions have been employed
\begin{align}
X(Q_1,Q_2,t) = & \ \frac{(1-t^2)^{-1/2}}{Q_1Q_2}\arctan\left(\frac{z\sqrt{1-t^2}}{1-zt}\right),  \\
z = & \ \frac{Q_1Q_2}{4m_{\mu}^2}(1-R_{m_1})(1-R_{m_2}), \\
R_{m_i}= & \ \sqrt{1+4m_{\mu}^2/Q_i^2}.
\end{align}

\section{TFF models}
\label{app:b}

\subsection{Regge model}

In this appendix we discuss in detail the Regge model introduced in section~\ref{sec:ca} and compare it with an approximation to it based on a finite set of resonances.  Our goal is to show how an approximation built from a finite set of resonances converges to the original model,  which contains on turn an infinite set of them. Several strategies are explored.
The Regge model in question is taken from Refs.~\cite{RuizArriola:2006jge,Arriola:2010aq} and reads
\begin{multline}
\label{eq:ReggeDV}
F^{\textrm{Regge}}_{\pi^0\gamma^*\gamma^*}(Q_1^2,Q_2^2) = \frac{aF_{\pi^0\gamma\gamma}}{Q_1^2-Q_2^2} \\ \times
    \frac{\left[ \psi^{(0)}\left(\frac{M^2+Q_1^2}{a}\right) -\psi^{(0)}\left(\frac{M^2+Q_2^2}{a}\right) \right]}{\psi^{(1)}\left(\frac{M^2}{a}\right)},
\end{multline}
where $\psi^{(n)}(z) = \partial^{n}\ln\Gamma(z)$ is the polygamma function, the parameter $a$ is the string tension (which is fixed to $1.3$~GeV$^2$ based on the study from Ref.~\cite{Masjuan:2012gc}) and $M=0.708$~GeV is chosen to match the slope parameter $b_{\pi}$~\cite{Masjuan:2012wy}. At high energies it behaves as $Q^{-2}\ln Q^2$ for the single virtual case~\cite{RuizArriola:2006jge}, whereas it behaves as $aF_{\pi^0\gamma\gamma}/(\psi^{(1)}(M^2/a))Q^{-2}$ for $Q_1^2=Q_2^2\equiv Q^2$ when $Q^2\to\infty$.

The Regge model can be expressed as well as as an {\textit{infinite}} sum over the resonances within the Regge trajectory weighted by their correspondent residues,
\begin{multline}
  F^{\textrm{Regge}}_{\pi^0\gamma^*\gamma^*}(Q_1^2,Q_2^2) = \frac{F_{\pi^0\gamma^*\gamma^*}}{\psi^{(1)}(M^2/a)} \\ \times \sum_{m=0}^{\infty} \frac{a^2}{(Q_1^2+ (M^2+ma))(Q_2^2+ (M^2+ma))}. \label{eq:ReggeSum}
\end{multline}

A resonance approach to this model will consist in retaining a finite number of resonances, 
achieving an increased precision as soon as more terms in the sum are included, assuming of course that the model parameters (masses and residues) are known. Such an approximation has, however, a slow rate of convergence, which is well understood from Pad\'e theory~\cite{Masjuan:2007ay}. To illustrate this, we perform a numerical test and show in the first row in Table~\ref{tab:regg} (called Res) what would be obtained for $a_{\mu}^{\textrm{HLbL};\pi^0}$ if truncating the sum in Eq.~\eqref{eq:ReggeSum} for a finite number of resonances $n$ (e.g., up to $m=n-1$ in Eq.~\eqref{eq:ReggeSum}). 
\begin{table}[h]
\centering 
\begin{tabular}{ccccc} \toprule
 $n$ & $1$ & $2$ & $3$ & $4$  \\ \hline
Res & $38.1$ & $47.1$ & $50.8$ & $52.8$  \\
Norm & $50.8$ & $57.0$ & $58.4$ & $59.1$  \\  
Der & $50.8$ & $57.8$ & $59.4$ & $59.9$  \\  \hline
Fit & $55.9$ & $67.2$ & $58.3$ & $65.4$  \\  
Fit$^{\textrm{OPE}}$ & $-$ & $63.3$ & $58.0$ & $61.5$  \\    \hline
Exact  & \multicolumn{4}{c}{$60.7$}  \\ \botrule
\end{tabular}
\caption{The $a_{\mu}^{\textrm{HLbL};\pi^0}$ result ($10^{-11}$ units) from different resonance-like approaches employed to approximate the Regge model which include up to $n$ resonances in Eq.~\eqref{eq:ReggeSum}. The exact result to compare with is represented in the last row. Find details in the text.\label{tab:regg}}
\end{table}
The slow asymptotic convergence is attributed to the fact that not even the TFF at the origin, $Q_1^2=Q_2^2=0$, is precisely reproduced. Therefore, to improve on that, we do not use the residue of the heavier resonance in the truncation; instead, we choose to fix such parameter to reproduce the TFF at the origin. As shown in the third row of Table~\ref{tab:regg} (called Norm) this strategy improves considerably on convergence. Finally, we choose to match not one, but all the residues in the summation, to fulfill the low-energy expansion. The results are shown in the fourth row of Table~\ref{tab:regg} (called Der) and yield the expected improvement on convergence. 

Besides, as customary in resonance approaches, one can fit the residues to a set of pseudodata instead of matching them to the low-energy expansion of the model. Performing such fit using the same points as for that in Table~\ref{tab:model} Fit$^{\textrm{OPE}}$ row, we obtain the results of fifth row in Table~\ref{tab:regg} (called Fit), which show an irregular convergence, if it converges at all. At this point, one could blame the incorrect behavior of the resulting approximant when both $Q_1^2=Q_2^2\equiv Q^2\to\infty$, that behaves as $Q^{-4}$ instead of $Q^{-2}$. Including additional terms to fulfill this behavior, we obtain the results in the sixth row in Table~\ref{tab:regg} (called Fit$^{\textrm{OPE}}$), which show an improved convergence, but still not such a good convergence as CAs to the same model with the same pseudodata fit (cf. Table~\ref{tab:model}). This appendix illustrates the potential systematic errors when the poles in rational approximants are fixed in advance.

\subsection{Logarithmic model}

Besides the Regge model, we introduce the logarithmic one employed as well in Section~\ref{sec:ca}. The latter is inspired in flat 
distribution amplitudes models as introduced in Ref.~\cite{Radyushkin:2009zg}, and is extended to the double-virtual case as follows
\begin{align}
\label{eq:logflat}
    F^{\textrm{log}}_{\pi^0\gamma^*\gamma^*}(Q_1^2,Q_2^2) &= \frac{F_{P\gamma\gamma}}{M^2} \int_0^1 dx \frac{1}{xQ_1^2 + (1-x)Q_2^2 + M^2} \nonumber \\
    &= \frac{F_{P\gamma\gamma}M^2}{Q_1^2-Q_2^2}\ln\left( \frac{1+Q_1^2/M^2}{1+Q_2^2/M^2} \right),
\end{align}
where $M=0.530$~GeV is chosen again to reproduce $b_{\pi}$~\cite{Masjuan:2012gc}. Again, it is straightforward to see its large $Q^2$-behavior $Q^{-2}\ln Q^2$ for a single-virtual photon; for equal virtualities, $F_{P\gamma^*\gamma^*}^{\textrm{log}}(Q^2,Q^2) = F_{P\gamma\gamma}M^2(M^2+Q^2)^{-1}$ and, as a consequence $\lim_{Q^2\to\infty}F_{P\gamma^*\gamma^*}^{\textrm{log}}(Q^2,Q^2)=F_{P\gamma\gamma}M^2Q^{-2}$.

\section{$P\to\bar{\ell}\ell$ decays impact}
\label{app:pll}

A further possibility to test the TFF double-virtual behavior is given by the $P\to\bar{\ell}\ell$ decays. Whereas these decays could never be directly employed to extract the TFF $q^2$-dependence, as it happens with $P\to\bar{\ell}\ell\bar{\ell}'\ell'$ decays, they offer an indirect probe in terms of a loop integral over the double-virtual TFF~\cite{Masjuan:2015lca}. The proposal to use these decays as a constraint for $a_{\mu}^{\textrm{HLbL;}P}$ was considered for the first time in Ref.~\cite{Hayakawa:1997rq}, but, to our best knowledge, it has not been seriously considered so far. 

In Refs.~\cite{Masjuan:2015lca,Masjuan:2015cjl}, we performed a detailed and careful study of these decays employing our method of CAs. So far, only the $\pi^0\to e^+e^-$~\cite{Abouzaid:2006kk} and $\eta\to\mu^+\mu^-$~\cite{Abegg:1994wx} have been measured. For the $\pi^0$, we found its experimental value $2\sigma$ away from our CAs prediction~\cite{Masjuan:2015lca}. Particularly, we found that in order to reproduce the experimental result would require $\delta_{\pi}^2\gtrsim 10~\textrm{GeV}^2$ in Eq.~\eqref{eq:ope} and $a_{\pi;1,1}< -4b_{\pi}^2$ in Eq.~\eqref{eq:TFFseries}. This would imply large corrections to the leading OPE behavior, a result far from theoretical expectations. In any case, taking these values for reconstructing the TFF\footnote{Here we assume the absence of new-physics effects.}, we would obtain that $a_{\mu}^{\textrm{HLbL;}\pi}$ would shift down to around $36(7)\times10^{-11}$~\cite{Masjuan:2015lca}. Of course, this result calls for a new experimental determination of the $\pi^0\to e^+e^-$ branching ratio, which could be possible at the NA62 experiment. Note that a first measurement of the double-virtual TFF at BES III would discard the $\pi^0$ TFF as the explanation for the measured branching ratio. Looking forward into the future, we study here the precision that such experiment would require in order to provide a valuable constraint to our study; this is, to improve our current $1.89b_{\pi}^2 \leq a_{\pi;1,1} \leq 2.10b_{\pi}^2$ range. We find that, due to the narrow range for $a_{\pi;1,1}$, a precision of $0.1\%$ on the BR would be required. Still, even if this precision may be out of experimental reach, a new experiment that would shed light on the  nature of the current deviation is highly desirable.

For the $\eta$ case, the sensitivity to the $a_{\eta;1,1}$ LEP is even lower. Even if an accurate description requires a proper double-virtual implementation, after implementing the parameters in Eq.~\eqref{eq:ope}, there is not much sensitivity to the $a_{\eta;1,1}$ parameter. Particularly, a precision below $0.1\%$ in the branching ratio measurement would be required to discern values within our range. Unfortunately, there are additional effects which are of further relevance at this precision, such as the implementation within our framework of the $\pi\pi$ cut, which would demand a more refined theoretical study~\cite{Masjuan:2015cjl}. In any case, a $2\%$ accuracy would be interesting already in order to corroborate (or falsify) our predictions, which may provide the only (indirect) experimental test of our TFF description in the near future. At present, the current precision is of $14\%$ and the central value is $1.4\sigma$ above our prediction. Similar comments apply for the $\eta'$ decays too, which however have never been  observed, being upper bounds the only available information so far~\cite{Akhmetshin:2014hxv,Achasov:2015mek}. It would be therefore an interesting possibility in the future to access $\eta(\eta')\to\mu^+\mu^-$ decays at LHCb~\cite{Huong:2016gob}, as well as the possibility to measure the $e^+e^-\to\eta'$ process at KLOE-II~\cite{Solodov}.

\section{The light-quark TFF}
\label{app:lq}

It is widely assumed that the smallness of the OZI rule violation would allow to express the $\eta$ and $\eta'$ TFFs in terms of the light- and strange-quark TFFs~\cite{Feldmann:1999uf}. In addition, it is believed that the former would resemble the $\pi^0$ one except for a $5/3$ charge factor. Under the validity of these assumptions, the knowledge of the  $\eta-\eta'$ mixing provides thereby  a cross-check for the $\pi^0$ TFF and, consequently, $a_{\mu}^{\textrm{HLbL};\pi^0}$. 

In our analyses from Refs.~\cite{Escribano:2013kba,Escribano:2015nra,Escribano:2015yup}, we could determine the mixing parameters, obtaining small violations of the OZI rule and confirming thereby the accuracy of such an approximation\footnote{From these studies we obtain the mixing angles $\phi_q\simeq\phi_s\simeq39.5^{\circ}$.}. From these results and those for $b_{\eta^{(\prime)}}, c_{\eta^{(\prime)}}, d_{\eta^{(\prime)}}$ and $\eta^{(\prime)}_{\infty}$ (cf. Table~\ref{tab:input}), we obtain a light-quark $\eta_q$ TFF which resembles the $\pi^0$ in the SL region, and seems to support Belle tendency~\cite{Uehara:2012ag} against the \babar~\cite{Aubert:2009mc} one at high energies, see Fig.~\ref{fig:lq}. For completeness we also show the strange-quark TFF understood as the one for a pure $s\bar{s}$ $\eta_s$ state. Fig~\ref{fig:lq} also shows the experimental data for the $\pi^0$ TFF from Refs.~\cite{Agashe:2014kda,Behrend:1990sr,Gronberg:1997fj,Aubert:2009mc,Uehara:2012ag,BABAR:2011ad,Acciarri:1997yx,BABAR:2011ad} (blue circles) and the experimental data for the $\eta$ and $\eta'$ TFFs rotated to the flavor light-strange-quark basis. The resulting parameters from the $\eta_q$ TFF read 
\begin{figure}[t]
\includegraphics[width=0.9\linewidth]{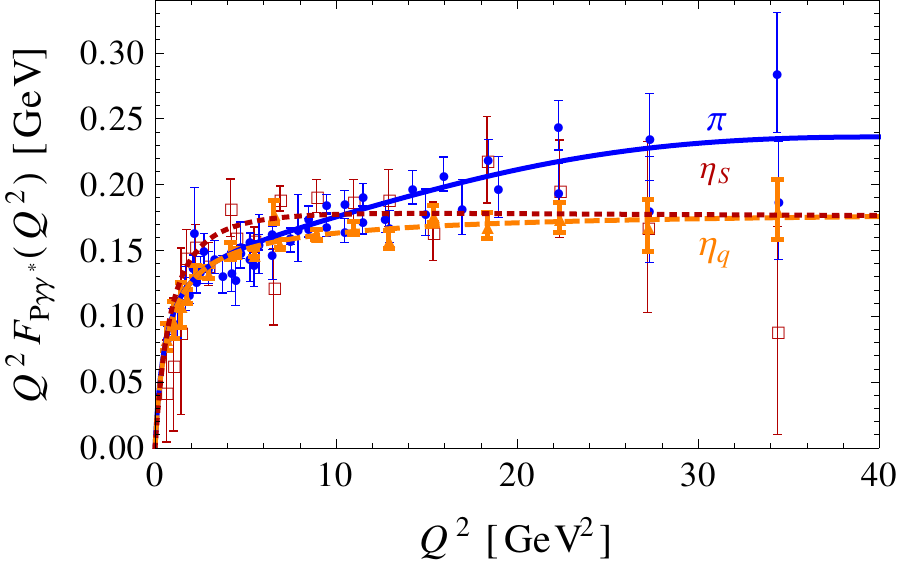}
\caption{Comparison of the $\pi^0$ TFF (blue solid) against light- (orange-dashed) and strange- (red-dotted) quark TFF. The experimental data from Refs.~\cite{Agashe:2014kda,Behrend:1990sr,Gronberg:1997fj,Aubert:2009mc,Uehara:2012ag,BABAR:2011ad,Acciarri:1997yx,BABAR:2011ad} is included as well as blue circles, orange triangles and empty-red squares for the $\pi^0$, light-quark and strange-quark TFFs, respectively (the latter are obtained after rotating the $\eta$ and $\eta'$ TFFs data).\label{fig:lq}}
\end{figure}
\begin{align} \nonumber
(3/5)F_{\eta_q\gamma\gamma} &=  0.2579(32) \textrm{GeV}^{-1},\\\nonumber
b_{\eta_q}m_{\eta_q}^{-2} &=  1.66(2)\textrm{GeV}^{-2}, \\\nonumber
c_{\eta_q}m_{\eta_q}^{-4} &=  2.87(8)\textrm{GeV}^{-4}, \\\nonumber
d_{\eta_q}m_{\eta_q}^{-6} &=  5.05(37)\textrm{GeV}^{-6}, \\\nonumber
\eta_{q\infty} &=  0.180(6)\textrm{GeV} .
\end{align}
Indeed, every parameter is compatible with those of the $\pi^0$ TFF except for the normalization. 

Taking the assumption that $F_{\pi^0\gamma^*\gamma^*}(Q_1^2,Q_2^2) = (3/5)F_{\eta_q\gamma^*\gamma^*}(Q_1^2,Q_2^2)$, we can provide an alternative determination for $a_{\mu}^{\textrm{HLbL;}\pi}$. This is found in the second row of Table~\ref{tab:lq}, called \textit{l.q.} A non-negligible shift appears with respect to our $\pi^0$ results in Table~\ref{tab:gm2}. However, this is to be expected given their different normalizations, which in turn represent the most relevant parameter in the calculation. Normalizing the light-quark TFF to match the $\pi^0$ one, an excellent agreement is found as shown in the third row of Table~\ref{tab:lq}, called \textit{l.q. norm}.
\begin{table}[t]
\begin{tabular}{cccc} \toprule
	$a_{\mu}^{\textrm{HLbL;}l.q.}$  & $C^0_1$  & $C^1_2[a_{P;1,1}^{\textrm{min}}]$ & $C^1_2[a_{P;1,1}^{\textrm{max}}]$ \\ \hline
	$l.q.$   & $60.4(1.5)(0.5)[1.6]$  & $57.2(1.8)[1.8]$ & $57.3(1.4)(1.0)[1.8]$ \\ 
	$l.q. \ norm$  & $67.4(1.7)(0.5)[1.8]$  & $63.8(2.0)[2.0]$ & $63.9(1.6)(1.1)[1.9]$ \\ \botrule
\end{tabular}
\caption{The analog results to those for $a_{\mu}^{\textrm{HLbL;}\pi^0}$ ($10^{-11}$ units) in Table~\ref{tab:gm2}, but employing the light-quark TFF. See details in the text.\label{tab:lq}}
\end{table}  
The errors identification is analog to that in Table~\ref{tab:gm2}. The values obtained in this exercise reassess our main results in Table~\ref{tab:gm2} and supports our statement that future experimental data at high SL energies would not change much the $a_{\mu}^{\textrm{HLbL};\pi^0}$ central value, whereas they will provide a preciser determination for the LEPs in Eq.~\eqref{eq:TFFseries}.

\section{Beyond pole contribution}
\label{app:beyondpole}

There is at present a debate on how to deal with the HLbL tensor high-energy behavior dictated by the OPE (cf. discussion in \cite{Jegerlehner:2009ry} with respect to \cite{Melnikov:2003xd} and the summary talk by Vainshtein in \cite{Proceedings:2016bng}). Whereas we do not want to enter this debate, in this appendix we discuss how our approach could be used by both approaches~\cite{Melnikov:2003xd,Jegerlehner:2009ry}. The first approach~\cite{Melnikov:2003xd} proposes to modify the $\pi^0$-pole contribution to $a_\mu$ as such that certain OPE constraint to the $\langle VVVV \rangle$ Green's function is satisfied. Its modification results on setting the external TFF (the gray blob connected to the external photon in Fig.~\ref{fig:pspole}) to a constant one. Following such prescription and using our description for the pseudoscalar TFFs, we obtain the results for $a_{\mu}^{\textrm{HLbL;}P}$ shown in Table~\ref{tab:mv}.
\begin{table}[h]
\begin{tabular}{cccc} \toprule
	$a_{\mu}^{\textrm{HLbL;}P}$  & $C^0_1$  & $C^1_2[a_{P;1,1}^{\textrm{min}}]$ & $C^1_2[a_{P;1,1}^{\textrm{max}}]$ \\ \hline
	$\pi^0$  & $84.9(1.8)(2.6)[3.2]$  & $82.8(1.7)[1.7]$ & $80.9(1.3)(0.5)[1.4]$ \\ 
	$\eta$  & $29.1(1.0)(0.3)[1.0]$  & $27.3(1.4)[1.4]$ & $26.9(1.5)(1.0)[1.8]$ \\ 
	$\eta'$  & $30.4(1.0)(0.5)[1.1]$  & $26.8(1.1)[1.1]$ & $25.8(0.7)(0.9)[1.1]$ \\ \hline
	Total  & $144.4[3.5]$  & $136.9[2.5]$ & $133.6[2.5]$ \\ \botrule
\end{tabular}
\caption{The results for $a_{\mu}^{\textrm{HLbL;}P}$ in units of $10^{-11}$ according to the procedure in Ref.~\cite{Melnikov:2003xd}. The errors and labeling are identical to those in Table~\ref{tab:gm2}.\label{tab:mv}}
\end{table}  
The larger errors obtained now are proportional to the larger central values with, essentially, the same proportionality than our main result. Accounting for the systematic error as we did in Section~\ref{sec:results}, we would obtain 
\begin{equation}\label{bpole}
a_{\mu}^{\textrm{HLbL}; P} = 135(11)\times10^{-11},
\end{equation}
to be compared with the result from Ref.~\cite{Melnikov:2003xd} $a_{\mu}^{\textrm{HLbL}} = (76.5+18+18)\times10^{-11}\to114\times10^{-11}$. This comparison illustrates again the potential large systematic errors --- beyond $10\%$ --- typical of resonance models. Actually, the result from~\cite{Melnikov:2003xd} was used in the \textit{Glasgow consensus} to obtain the reference value $10.5(2.6)$. If we would replace their pseudoscalar-pole contribution by our Eq.~\ref{bpole}, the final result would be
\begin{equation}\label{LbL}
a_{\mu}^{\textrm{HLbL}} = 126(25)\times10^{-11},
\end{equation}
\noindent
one sigma larger, and in better agreement with ballpark estimates.

The second approach~\cite{Jegerlehner:2009ry} provides instead a model for the pseudoscalar contribution (not only the lightest pseudoscalar poles), related to the $\langle VVP \rangle$ Green's function. It would be possible within our approach to reconstruct an analogous Green's function. In this scenario to satisfy all the constraints imposed by the OPE, one should start directly with the $C^1_2(Q_1^2,Q_2^2,(Q_1+Q_2)^2)$ since the $N=0$ is too limited. Then, we cannot provide a systematic error and check on convergence. Besides, further constraints on this $\langle VVP \rangle$ Green function would be desired. For these reasons, we decide not to give a value for such scenario. In any case, we remind that this approach, as well as the previous one, were inspired in the $\pi^0$ TFF model in Ref.~\cite{Knecht:2001qf} which, as said, entails non-accounted systematic errors.

\section{Stieltjes functions}
\label{app:stieltjes}

A function is said to be of the Stieltjes kind if it admits an integral representation~\cite{BakerMorris}
\begin{equation}\label{Stieltjes}
f(q^2) = \int_0^{1/R} \frac{d \phi(u)}{1-u q^2} \ ,
\end{equation}
where $\phi(u)$ is any bounded and nondecreasing function~\cite{BakerMorris}. To see that such is the case for the isovector contribution to the TFF in Refs.~\cite{Stollenwerk:2011zz,Hanhart:2013vba,Kubis:2015sga}, let $R=4m_{\pi}^2$, and define $d \phi(u) = {\rm const.}\times\frac{q^2}{\pi}\frac{\operatorname{Im}F(1/u)}{u}$; making the change of variables $u=1/s$, Eq.~(\ref{Stieltjes}) returns the once-substracted dispersive representation of the isovector contribution discussed in Ref.~\cite{Stollenwerk:2011zz}, and also exploited in Refs.~\cite{Hanhart:2013vba,Kubis:2015sga}, once $\operatorname{Im}F(s)= \sigma^3(s) P(s) |F_V(s)|^2$ is identified. Since $\sigma(s)=\sqrt{1-4m_{\pi}^2/s}$, $P(s)$ is a linear polynomial with positive slope and $F_V(s)$ the $\pi^{\pm}$ vector FF, then $\operatorname{Im}F(s)$  is a positive function, the requirement of $\phi(u)$ to be nondecreasing is fulfilled and the convergence of PAs to the TFF is guaranteed.\footnote{If the function $f(z)$ is a Stieltjes function, its $n^{th}$-subtracted version is a Stieltjes function as well~\cite{BakerMorris}.}

\section{Dispersion Relations}
\label{app:dr}

In this appendix we develop our statements concerning potential drawbacks of dispersive approaches for extending the TFF representation into the SL region beyond energies of the order of $1$~GeV. For this purpose, we employ a simplified approach inspired from Ref.~\cite{Hanhart:2013vba}. Specifically, we take the definition in Eq.~(17) of that reference for the once-subtracted dispersion relation for the $\eta$ TFF, while we adopt a simpler but reasonable description for the $\pi^{\pm}$ vector form factor based on Refs.~\cite{GomezDumm:2000fz,Dumm:2013zh}. 

The result obtained for the TL region (from $q^2=0$ up to $q^2=m_{\eta}^2$), accessible in the $\eta\to\gamma\bar{\ell}\ell$ Dalitz decays, is illustrated in the upper panel of Fig.~\ref{fig:DRSL}, and shows a nice agreement with existing data even though the approximations performed; 
\begin{figure}[b]
\includegraphics[width=0.8\linewidth]{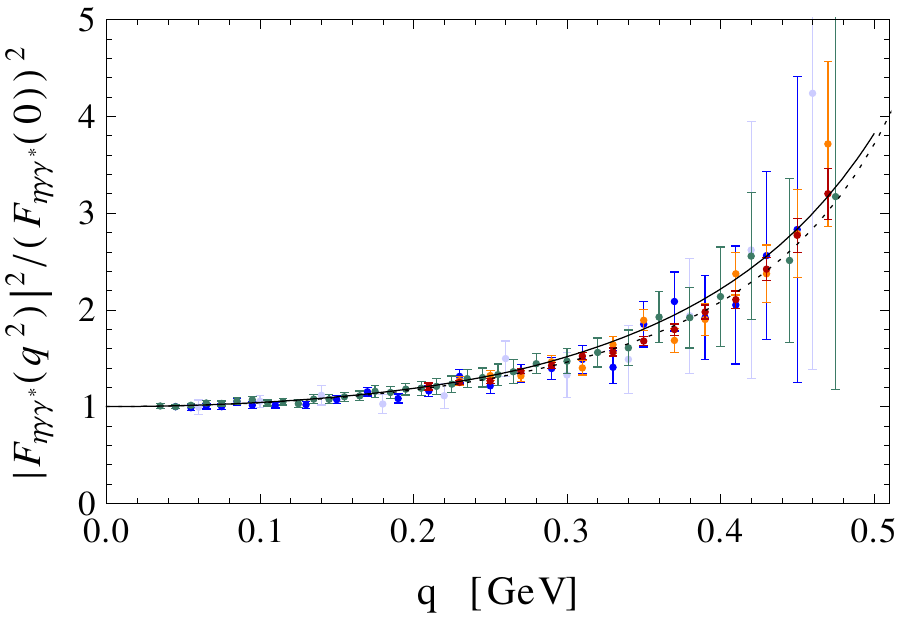}\\
\vspace{0.4cm}
\includegraphics[width=0.8\linewidth]{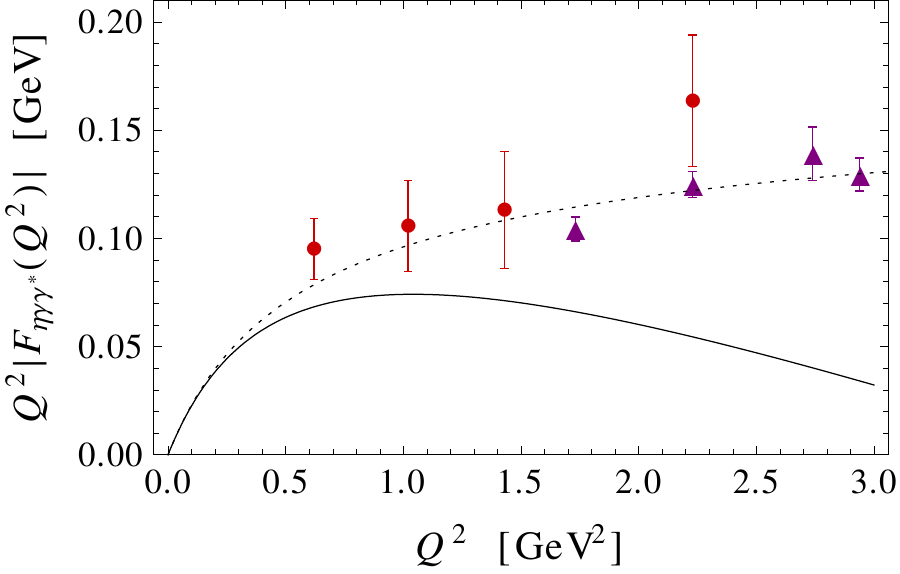}
\caption{Comparison of a dispersive representation for the $\eta$ TFF in the low-energy TL and SL regions (upper and lower figure respectively) to the available data from Refs.~\cite{Arnaldi:2009aa,Berghauser:2011zz,Aguar-Bartolome:2013vpw,Arnaldi:2016pzu,Adlarson:2016hpp} and Refs.~\cite{Behrend:1990sr,Gronberg:1997fj}, respectively. We also show our parameterization using CAs as black-dotted line.\label{fig:DRSL}}
\end{figure}
this nice overlap contrasts with the situation in the SL region at higher energies, which is illustrated in the lower panel of Fig.~\ref{fig:DRSL} and shows a clear deterioration above $1$~GeV. Of course, we stress that this is an oversimplified dispersive model and avoids for instance heavier resonances or inelasticities as done in Refs.~\cite{Kubis:2015sga,Xiao:2015uva}, but we hope it is enough to illustrate the expected features for large $Q^2$ values from dispersive representations, which are tightly related to the subtraction procedure. 

In this respect, it would be interesting to use the series expansion at $Q^2=0$ that dispersive approaches can provide --- given their reliability at low energies --- to supply further information for the rational approach adopted in this work. Such procedure would provide a possibility to extend dispersive approaches up to arbitrarily large $Q^2$-values in the SL or, at the very least, to implement their LEPs. 

\section{Input parameters}
\label{app:inputs}

In this appendix we quote the different inputs used in our calculation. They are collected in Table~\ref{tab:input} together with their original reference(s) and the data involved in the analysis for such determination. In addition, we quote in the last column those experiments which could improve the present values for the employed parameters. These parameters together with the definitions in Sections~\ref{sec:ppole} and~\ref{sec:ca} and Appendix~\ref{app:a} should allow the reader to reproduce all the results presented in this work. 
 
We note that we have taken the opportunity to incorporate the recently released data from Refs.~\cite{Arnaldi:2016pzu,Adlarson:2016hpp}\footnote{The experimental data from Ref.~\cite{Adlarson:2016hpp} supersedes the one used~\cite{Aguar-Bartolome:2013vpw} in our previous work~\cite{Escribano:2015vjz}, which we remove in consequence.} and Refs.~\cite{Adlarson:2016ykr,TheNA62:2016fhr}\footnote{Ref.~\cite{TheNA62:2016fhr} does not include the data points. Therefore, to include their analysis we used two different strategies: the first consists in generating a single point using their fit result, whereas the second consists in including their slope as a fitting parameter  and leads to analog results. We thank M.~Koval for discussions.} regarding the $\eta$ and $\pi^0$ TFFs. For the $\eta$ case, we obtain very similar results to those in Ref.~\cite{Escribano:2015vjz}, with the advantage of reaching the $P^3_3$ approximant and obtaining a reduced statistical uncertainty due to the improvement in the experimental precision. For the $\pi^0$ case, the changes in central values are due to the updated $\Gamma(\pi^0\to\gamma\gamma)$ value from Ref.~\cite{Agashe:2014kda} with respect to that used in Ref.~\cite{Masjuan:2012wy}, the inclusion of systematic uncertainties for \cite{Behrend:1990sr} and, mildly, the data from Refs.~\cite{Adlarson:2016ykr,TheNA62:2016fhr}. The improvement on systematics is low as a consequence of the limited and low TL $q^2$-range and we adopt those in~\cite{Masjuan:2012wy}. As an example, for the slope we obtain $b_{\pi}=0.0336(29), 0.0321(13)$ and $0.0315(15)$ for the $P^N_1$, $P^N_N$ and $P^N_N$ sequence with the Brodsky-Lepage asymptotic behavior built-in (we reach up to the $N=6,2$ and $3$, element respectively). As usual, we take the average as our final result~\cite{Masjuan:2012wy,Escribano:2013kba,Escribano:2015vjz,Escribano:2015yup}, which is quoted in Table~\ref{tab:input} and includes the systematic error, which has been combined in quadrature with the statistical one.
\begin{table}[t]
\centering\footnotesize
\begin{tabular}{ccccc} \toprule
Input & Value & Refs. & Data & Future\\ \hline
$F_{\pi\gamma\gamma}$   & $0.2724(29)$ &    &  \cite{Agashe:2014kda}  &  \cite{Gasparian:2016oyl,Babusci:2011bg}\\
$b_{\pi}$ & $0.0321(19)$ & \cite{Masjuan:2012wy}    & \cite{Behrend:1990sr,Gronberg:1997fj,Aubert:2009mc,Uehara:2012ag,Adlarson:2016ykr,TheNA62:2016fhr}  &  \cite{Babusci:2011bg,Adlarson:2014hka,Gan:2015nyc} \\
$c_{\pi}$ & $0.00104(22)$ & \cite{Masjuan:2012wy}   & \cite{Behrend:1990sr,Gronberg:1997fj,Aubert:2009mc,Uehara:2012ag,Adlarson:2016ykr,TheNA62:2016fhr}   & \cite{Babusci:2011bg,Adlarson:2014hka,Gan:2015nyc} \\
$d_{\pi}$ &      &     &   &  \cite{Babusci:2011bg,Adlarson:2014hka,Gan:2015nyc} \\
$a_{\pi;1,1}$ &      &    &   &  \cite{Adlarson:2014hka} \\
$\pi_{\infty}$ & $2F_{\pi}$ & \cite{Agashe:2014kda}   &    &  \\
$\delta_{\pi}^2$ & $0.20(2)$ & \cite{Novikov:1983jt,Jegerlehner:2009ry}   &   &  \\ \hline
$F_{\eta\gamma\gamma}$   & $0.2738(47)$ &     & \cite{Agashe:2014kda}   &  \cite{Adlarson:2014hka} \\
$b_{\eta}$ & $0.572(8)$ & \cite{Escribano:2015vjz,Escribano:2013kba}  &  \cite{Behrend:1990sr,Gronberg:1997fj,BABAR:2011ad,Arnaldi:2009aa,Berghauser:2011zz,Aguar-Bartolome:2013vpw,Arnaldi:2016pzu,Adlarson:2016hpp}  &  \cite{Adlarson:2014hka,Gan:2015nyc,Gan:2015nyc} \\
$c_{\eta}$ & $0.333(9)$ & \cite{Escribano:2015vjz,Escribano:2013kba}  & \cite{Behrend:1990sr,Gronberg:1997fj,BABAR:2011ad,Arnaldi:2009aa,Berghauser:2011zz,Aguar-Bartolome:2013vpw,Arnaldi:2016pzu,Adlarson:2016hpp}  &  \cite{Adlarson:2014hka,Gan:2015nyc} \\
$d_{\eta}$ &  $0.195(20)$ & \cite{Escribano:2015vjz,Escribano:2013kba}  & \cite{Behrend:1990sr,Gronberg:1997fj,BABAR:2011ad,Arnaldi:2009aa,Berghauser:2011zz,Aguar-Bartolome:2013vpw,Arnaldi:2016pzu,Adlarson:2016hpp}  &  \cite{Adlarson:2014hka,Gan:2015nyc} \\
$\eta_{\infty}$ & $0.180(12)$ & \cite{Escribano:2015vjz,Escribano:2013kba}   & \cite{Behrend:1990sr,Gronberg:1997fj,BABAR:2011ad,Arnaldi:2009aa,Berghauser:2011zz,Aguar-Bartolome:2013vpw,Arnaldi:2016pzu,Adlarson:2016hpp}   & \cite{Adlarson:2014hka,Gan:2015nyc} \\ \hline
$F_{\eta'\gamma\gamma}$    & $0.3437(55)$ &  & \cite{Agashe:2014kda} & \cite{Dudek:2012vr,Adlarson:2014hka} \\
$b_{\eta'}$ & $1.31(3)$ & \cite{Escribano:2015yup,Escribano:2013kba}   & \cite{Behrend:1990sr,Gronberg:1997fj,Acciarri:1997yx,BABAR:2011ad,Ablikim:2015wnx}   &  \cite{Dudek:2012vr,Adlarson:2014hka}\\
$c_{\eta'}$  & $1.74(9)$ & \cite{Escribano:2015yup,Escribano:2013kba}   & \cite{Behrend:1990sr,Gronberg:1997fj,Acciarri:1997yx,BABAR:2011ad,Ablikim:2015wnx}   & \cite{Dudek:2012vr,Adlarson:2014hka} \\
$d_{\eta'}$   & $2.30(22)$ & \cite{Escribano:2015yup,Escribano:2013kba}   &  \cite{Behrend:1990sr,Gronberg:1997fj,Acciarri:1997yx,BABAR:2011ad,Ablikim:2015wnx}  & \cite{Dudek:2012vr,Adlarson:2014hka} \\
$\eta'_{\infty}$  & $0.255(4)$ & \cite{Escribano:2015yup,Escribano:2013kba}   & \cite{Behrend:1990sr,Gronberg:1997fj,Acciarri:1997yx,BABAR:2011ad,Ablikim:2015wnx}  & \cite{Dudek:2012vr,Adlarson:2014hka}  \\ \botrule
\end{tabular}
\caption{The values for the different parameters employed in calculating $a_{\mu}^{\textrm{HLbL;}P}$ (second column) and the reference where they were extracted from (third column). $F_{P\gamma\gamma}$, $P_{\infty}$ and $\delta_{\pi}$ are expressed in GeV$^{-1}$, GeV and GeV$^{-2}$ units, respectively (additional parameters are dimensionless). The employed and expected future data are given in the fourth and fifth column, respectively.
\label{tab:input}}
\end{table}

In addition, we give the $Q^2F_{\pi^0\gamma^*\gamma}(Q^2)$ parameterization obtained from our the highest element reached within the $P^N_N$ sequence with the BL behavior built-in ($N=3$), 
\begin{equation}
\label{eq:ffpars}
 P^3_3(x) = x \frac{t_0 + t_1 x + t_2x^2}{1+r_1x + r_2x^2 + r_3x^3},  
\end{equation}
which coefficients are shown in Table~\ref{tab:ffpars} (only the central values are given).
\begin{table}[h]
\begin{tabular}{cccccc}\toprule
 $t_0$ & $t_1$ & $t_2$  &  $r_1$ & $r_2$ & $r_3$ \\ \hline
 $0.276$ & $-0.00024$ & $0.00024$ & $1.729$ & $-0.0285$ & $0.0013$ \\ \botrule
\end{tabular}
\caption{Parameters of the TFF parameterization Eq.~\eqref{eq:ffpars} obatined from our fitting procedure.\label{tab:ffpars}}
\end{table}

\newpage

\bibliography{HLbL_v7}

\end{document}